\documentclass[sn-mathphys-num]{sn-jnl}% Math and Physical Sciences Numbered Reference Style 
\usepackage{graphicx}%
\usepackage{multirow}%
\usepackage{amsmath,amssymb,amsfonts}%
\usepackage{amsthm}%
\usepackage{mathrsfs}%
\usepackage[tbtags]{mathtools}
\usepackage[title]{appendix}%
\usepackage{xcolor}%
\usepackage{textcomp}%
\usepackage{manyfoot}%
\usepackage{booktabs}%
\usepackage{algorithm}%
\usepackage{algorithmicx}%
\usepackage{algpseudocode}%
\usepackage{listings}%
\usepackage{latexsym}%
\usepackage{bbold}%
\usepackage{units}%
\usepackage{nicefrac}%
\usepackage{float}%
\theoremstyle{thmstyleone}%
\theoremstyle{thmstyletwo}%

\theoremstyle{thmstylethree}%

\newcommand{\be}{\begin{equation}}
\newcommand{\ee}{\end{equation}}
\newcommand{\bea}{\begin{eqnarray}}
\newcommand{\eea}{\end{eqnarray}}

\newcommand{\cL}{\mathcal{L}}
\newcommand{\cO}{\mathcal{O}}

\newcommand{\Pint}{-\hspace{-2.5ex}\int}

\newcommand{\fs}{\; \; .}

\newcommand{\chpt}{$\chi$PT~}
\newcommand{\KL}{K\"all\'en-Lehmann~}

\newcommand{\re}{\mathrm{Re}}
\newcommand{\im}{\mathrm{Im}}

\newcommand{\GeV}{\,\text{GeV}}

\newcommand{\eg}{{\em e.g. }}
\newcommand{\ie}{{\em i.e. }}

\newcommand{\EL}{E\L{}}

\begin{document}

\title{A brief introduction to dispersive methods}


\author[1]{\fnm{Gilberto} \sur{Colangelo}}\email{gilberto@itp.unibe.ch}
\affil[1]{\orgdiv{Albert Einstein Center for Fundamental Physics, \\
    Institute for Theoretical Physics}, \orgname{University of Bern}, \orgaddress{\street{Sidlerstrasse 5}, \city{Bern}, \postcode{3012}, \country{Switzerland}}}


\abstract{These lectures aim to provide a basic introduction to dispersive
  methods and their modern applications to the phenomenology of the
  Standard Model at low energy. This approach exploits analyticity
  properties of Green functions and scattering amplitude, often combined
  with unitarity constraints. To find a logically coherent set of topics in
  this vast subject, I start with the two-point Green's function, show that
  this needs the three-point function as input which in turn needs the
  four-point function. The sequence stops here, just like these lectures,
  because the four-point function is related only to itself (if one ignores
  inelastic effects), I will discuss these dispersion relations both in the
  case of toy models, simple scalar theories, as well as in the
  phenomenologically relevant case of QCD. The two-point function of the
  electromagnetic current in QCD plays a role in the evaluation of the
  hadronic vacuum polarization contribution to the $g-2$ of leptons. The
  most important contribution to this two-point function is due to the
  two-pion intermediate state. To evaluate this one needs the
  electromagnetic form factor of the pion as input. The dispersion relation
  for the latter takes the form of an Omn\`es problem and the solution is
  given by the Omn\`es function, which can be expressed in terms of the
  phase-shifts for the $\pi \pi$ scattering amplitude. The dispersion
  relation for the $\pi \pi$ scattering amplitude takes the form of the
  so-called Roy equations.  In this case we encounter for the first time a
  left-hand cut and see that this is constrained by crossing symmetry. If
  one takes into account also the nonlinear constraints given by unitarity
  one ends up with the Roy equations, which take the form of coupled,
  non-linear integral equations. A brief discussion of their numerical
  solutions and a few selected applications concludes the lectures.}




\maketitle

\newpage

\tableofcontents

\vskip 0.5cm

\section{Introduction} 

Weinberg's books on Quantum Field Theory
(QFT)~\cite{Weinberg:1995mt,Weinberg:1996kr} have the declared goal to show
that QFT is the way it is because of the fundamental principles of
analyticity, unitarity, Lorentz symmetry and cluster decomposition (which
is essentially locality). We want these principles to be respected and
therefore must describe nature with the QFT formalism, at least in the
range of energies we have accessed so far. Whatever form a more fundamental
theory may take at very high energy, at low energies it will always look
like a QFT. This justifies the view of QFT as a low-energy tool and of any
specific QFT as an effective field theory (EFT), only valid in a certain
energy range.

If one adopts this point of view a natural question is whether one can
derive consequences of these principles without explicitly relying on a
specific Lagrangian. This is particularly relevant for QFTs in their
non-perturbative regime, because in such cases the Lagrangian is not
particularly useful, no matter how simple it looks on paper, unless one
resorts to numerical methods like lattice---which implies discretizing
spacetime, carefully choosing the discretized version of the Lagrangian and
letting supercomputers do the work.

In these short lecture notes I will provide an introduction to methods
which implement the constraints of analyticity and unitarity and go under
the name of dispersion relations (DR). These have been developed and
extensively used in the sixties and seventies, mainly as an alternative to
QFT to describe the strong interactions. After the discovery of asymptotic
freedom and the formulation of QCD they have been mostly abandoned, but are
experiencing a renaissance lately, both as a tool to tackle strong
interactions at low energy (mostly in combination with chiral perturbation
theory, \chpt \hskip -0.1cm), as well as in an attempt to develop a more general
theoretical tool in the program which goes under the name of S-matrix
bootstrap. The focus of these lectures will be on the use of DR in
connection with the phenomenology of strong interactions at low energy and
in combination with \chpt, but I will also emphasize the complementarity
and the synergies with the lattice approach.

The structure of the lecture is as follows: I will start with a discussion
of the two-point function in a simple scalar theory and derive the \KL
representation. I will then show that if we introduce an additional strong
coupling in this theory and make it non-perturbative, the \KL
representation becomes even more useful as it relates the two-point
function to a form factor. This is essentially what is done in QCD+QED and
forms the basis of the dispersive evaluation of the HVP contribution to the
muon $g-2$.

I will then discuss the DR for pion form factors and their solution in
terms of the Omn\`es function, which expresses them in terms of their
phases only (and their zeros, if they have any). In the elastic region the
phase of the form factors coincides with the $\pi \pi$ scattering phase
shift of the same quantum numbers, which leads us to the next topic: the
$\pi \pi$ scattering amplitude. Due to the presence of crossed channels and
of crossing symmetry, the DR for a scattering amplitude is more complicated
than that for a form factor. I will derive that first in the case of a
simple $\lambda \phi^4$ theory, and then extend it to the realistic case of
pions, which have in addition the isospin degree of freedom, which adds a
further layer of complexity. The DR equations for pions have been first
derived by S.M.~Roy and go under the name of Roy equations.  I will discuss
in detail their numerical solutions, as well as the very precise results
which can be obtained by matching the Roy solutions to the chiral
representation in the subthreshold region. As a particularly striking
application of this formalism, I then discuss the determination of the
$\sigma$-resonance parameters, which is the last topic of these lecture
notes. In my presentation I will only consider the isospin limit of QCD
even though, at the present level of precision in the comparison with
experiments, isospin-breaking effects are relevant when discussing the
phenomenology, as I will briefly discuss with the help of one example.
Given space constraints I am only able to scratch the surface of this vast
subject here, but the final remarks offer to the interested reader some
perspective on further applications.

\section{Two-point function: from a scalar theory to QCD}
We start by considering the two-point function and to fix the framework we
will discuss this in a scalar theory containing a real ($\varphi$) and a complex
($\phi$) scalar (as done in Coleman's book on QFT~\cite{Coleman:2018mew}):
\be
\cL=\frac{1}{2}\partial_\mu \varphi \partial^\mu \varphi - \frac{1}{2} m^2
\varphi^2 + \partial_\mu \phi^* \partial^\mu \phi - M^2 |\phi|^2 - g \varphi
\phi^* \phi \; .
\label{eq:LagrY}
\ee
The Fourier transform of the two-point function of the real scalar
(assuming that it satisfies the renormalization condition $\langle 0|
\varphi(0) | 0 \rangle=0$):
\begin{align}
\tilde G(p,p')& = \int d^4x d^4 y e^{i(px +p'y)} \langle 0 | T \varphi(x)
\varphi(y) | 0 \rangle  =: (2 \pi)^4 \delta^4(p+p') \tilde{D}(p^2) \; , 
\end{align}
satisfies the K\"all\'en-Lehmann representation:
\be
\tilde{D}(p^2)=\frac{i}{p^2-m^2+ i \varepsilon}+ i\int_{4 M^2}^\infty ds
\frac{\bar\rho(s)}{p^2-s+ i \varepsilon} 
\label{eq:KL}
\ee
where the spectral function $\bar\rho(s)$ is defined as
\be
\bar\rho(q^2) \theta(q^0) = \Sigma'_n (2\pi)^3 \delta^4(q-p_n) |\langle n|
\varphi(0)| 0 \rangle|^2 
\label{eq:barrho}
\ee
where the sum runs over all possible asymptotic states $|n \rangle$ in the
theory, $p_n$ is the corresponding momentum, and  the prime in the sum means
that the single-particle state created by the $\varphi$ field is excluded
from it. The spectral function can also be defined as
$\rho(s):=\delta(s-m^2)+\bar\rho(s)$, in which case Eq.\eqref{eq:KL} can be
written without the explicit pole term and $\bar\rho \to \rho$.

The function $-i \tilde{D}(p^2)$ is analytic everywhere on the complex
$p^2$-plane, with the exception of the pole at $m^2$ and the cut starting at
$4 M^2$.\footnote{I am assuming here that the theory is perturbative and
that the lowest-lying intermediate state is the two-complex-scalars
state. Otherwise, the cut could start below that threshold.}
Since $-i \tilde{D}(p^2)$ is real on the real axis for $p^2< 4 M^2$ (and
$p^2\neq m^2$), it has to satisfy the Schwarz reflection principle:
\be
-i \tilde{D}(z^*)= \left[ -i \tilde{D}(z) \right]^*
\ee
which implies that the discontinuity across the cut is given
by\footnote{Upon making use of the Sokhotski-Plemelj identity
$\lim_{\varepsilon \to 0}\frac{1}{x+i \varepsilon} = -i \pi \delta(x) +P
\frac{1}{x}$.}. 
\be
\mathrm{Disc} \left[ -i \tilde{D}(p^2) \right]=-2 \pi i \bar\rho(p^2) \; ,
\label{eq:DiscD}
\ee
where $\mathrm{Disc}[f(s)]\equiv \lim_{\epsilon \to 0}[f(s+i
\epsilon)-f(s-i \epsilon)]$.

If we introduce the self-energy $-i\tilde{\Pi}(p^2)$, \ie the sum of all
possible one-particle irreducible (1PI) contributions to the two-point
function, we can express $\tilde{D}(p^2)$ in terms of  $\tilde{\Pi}(p^2)$
as follows: 
\be
\tilde{D}(p^2) = \frac{i}{p^2-m^2-\tilde{\Pi}(p^2) + i \varepsilon} \; ,
\ee
after having resummed the geometric series of all possible insertions of
the self-energy. We define the renormalized self-energy
$\tilde{\Pi}_r(p^2)$ as satisfying standard renormalization conditions:
\be
\tilde{\Pi}_r(m^2)=0\;, \qquad \frac{\partial}{\partial p^2}
\tilde{\Pi}_r(p^2)_{|_{p^2=m^2}}=0 \; ,
\label{eq:RenC}
\ee
and correspondingly, the renormalized two-point function
$\tilde{D}_r(p^2)$. 

In perturbation theory it is easy to calculate the self-energy, and at
order $g^2$ one obtains:
\be
\tilde{\Pi}_{g^2}(p^2)= i g^2 \int \frac{d^4q}{(2 \pi)^4} \int_0^1 dx
\frac{1}{\left[q^2+p^2x(1-x)-M^2+i \varepsilon \right]^2} \; ,
\ee
and after imposing the renormalization conditions~\eqref{eq:RenC}
\be
\tilde{\Pi}_{g^2}(p^2)=\frac{g^2}{16 \pi^2} \! \int_0^1 \!\!\! dx \left[ 
\log \left(\frac{M^2-p^2 x(1-x) -i \varepsilon}{M^2-m^2 x(1-x)} \right)
+\frac{(p^2-m^2)x(1-x)}{M^2-m^2x(1-x)}\right]  \; . 
\label{eq:Pirexpl}
\ee 
For real values of its argument and $p^2 \leq 4 M^2$ the self-energy is
real and can be evaluated for $\varepsilon=0$. But for $p^2> 4 M^2$ it
becomes complex, with the imaginary part arising from the
logarithm and whose sign is fixed by the presence of the $i \varepsilon$ of
which we have kept track through the whole calculation. Remembering that,
for $x>0$, $\log(-x \pm i \varepsilon)= \log|x|\pm i \pi$ and after
identifying the region in which the argument of the log is negative, we
obtain (for $p^2>4M^2$):
\be 
\mathrm{Im} \tilde{\Pi}_{g^2}(p^2) = -\frac{g^2}{16 \pi^2} \int_{-
  \sigma(p^2)/2}^{\sigma(p^2)/2} \!\!\!\!\! dy \; \pi = -\frac{\alpha_g}{4}\sigma(p^2)
\label{eq:ImPi}
\ee
with
\be
\alpha_g := \frac{g^2}{4 \pi} \; \; \mbox{and} \; \; \; \sigma(s)=\sqrt{1-\frac{4 M^2}{s}} \; .
\ee

I have singled out the imaginary part from the loop calculation because
this is predicted specifically by the K\"all\'en-Lehmann representation,
through Eqs.~\eqref{eq:DiscD} and \eqref{eq:barrho}. To make this
completely explicit we need to calculate the matrix element of the
$\varphi$ field between the vacuum and the $2 \phi$-state:
\be
| \langle 2 \phi| \varphi(0)| 0 \rangle |^2 = \frac{g^2}{(p^2-m^2)^2}
\label{eq:2phivac}
\ee
which corresponds to the $\cO(g^2)$ contribution to the self-energy. To
complete the calculation we still need to evaluate the sum over all
possible such states, \ie integrate over the two-particles phase space:
\be
\int \frac{d^3p_1}{(2 \pi)^3 2 E_1} \int \frac{d^3p_2}{(2 \pi)^3 2 E_2} (2
\pi)^4 \delta^4(q -(p_1+p_2)) = \frac{1}{8 \pi} \sigma(q^2) \; .
\ee
Multiplying this by the matrix element~\eqref{eq:2phivac} and dividing by $2 \pi$ we
end up with
\be
\bar\rho(p^2) = \frac{\alpha_g}{4 \pi } \frac{\sigma(p^2)}{(p^2-m^2)^2} \;,
\ee
which confirms the result of the explicit calculation, Eq.~\eqref{eq:ImPi},
after expanding to $\cO(g^2)$ the relation $\mathrm{Im}\left[-i
  D(p^2)\right]=-\pi \bar\rho(p^2)=
\mathrm{Im}\tilde{\Pi}_r(p^2)/|p^2-m^2-\tilde{\Pi}_r(p^2)|^2 $. 

\begin{figure}[t]
\centering
\includegraphics[width=12cm, trim= 1cm 11cm 2cm 3cm, clip]{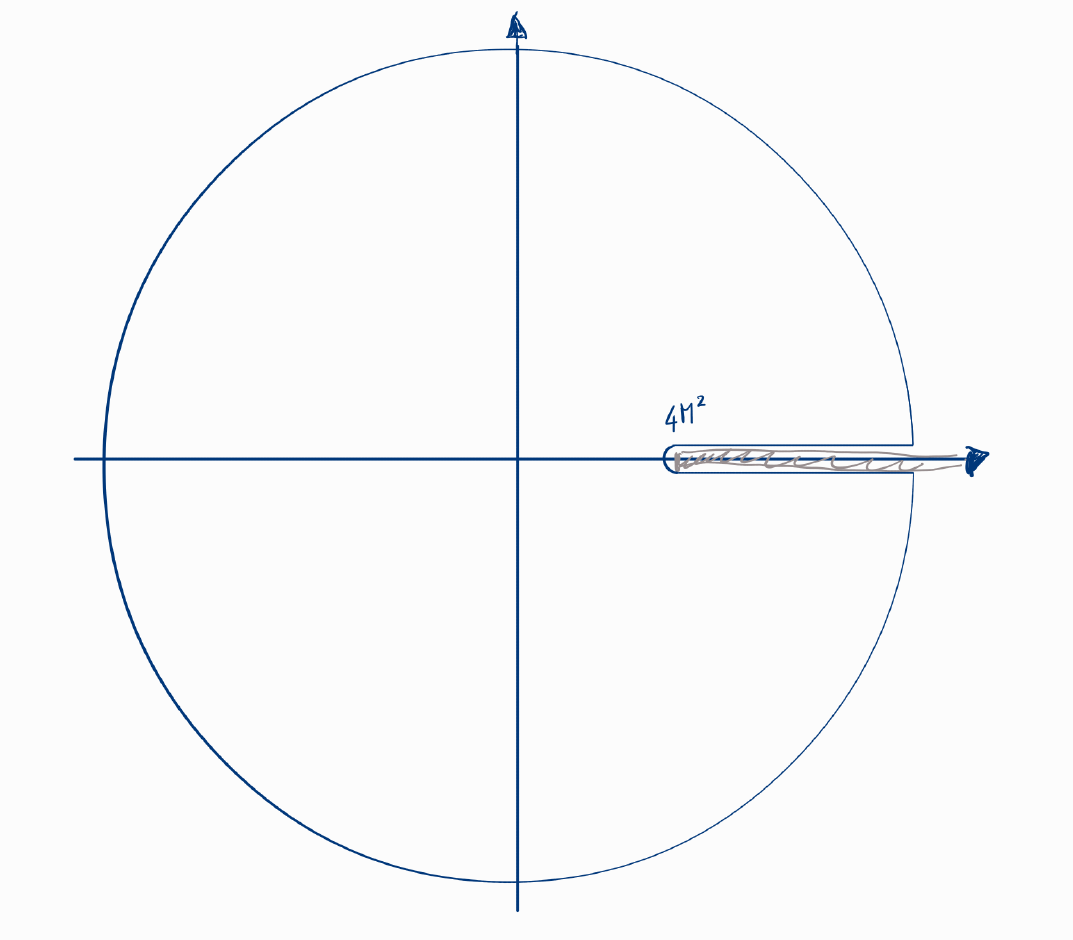}
\caption{Integration contour used in the derivation of the dispersion
  relation for a function with a right-hand cut starting at $4 M^2$.
\label{fig:contour}}
\end{figure}

But not only the imaginary part can be predicted from the
K\"all\'en-Lehmann representation. Since this incorporates also the
analytic properties of the two-point function $\tilde{D}$, which get
inherited by the self-energy $\tilde{\Pi}_r(p^2)$, one can use these to
express also the real part in terms of its discontinuity through a
dispersive integral. This step is the basis of all dispersion
relations, so it is useful to discuss it in detail here. Using Cauchy's
theorem it is easy to show that any complex function can be expressed in
terms of the following contour integral:
\be
f(s)= \frac{1}{2 \pi i} \oint_C dz \frac{f(z)}{z-s} \; .
\ee
provided the contour $C$ is inside the domain of analyticity of the
function $f$ and $s$ is inside the contour. In our case we have to deal
with a function which is analytic in the whole complex plane with the
exception of a cut on the real axis between $4 M^2$ and $\infty$, so any
contour of the form shown in Fig.~\ref{fig:contour} would work.
We can also send the radius of the circle to infinity and, if the function
considered vanishes for $|z| \to \infty$, the integral over the circle at
infinity will not contribute at all. In this case the function can be
expressed as the difference between the integral on the upper rim of
the cut and the one on the lower rim of the cut. In other words, as the
integral over the discontinuity:
\be
f(z) = \frac{1}{2 \pi i} \int_{4 M^2}^\infty ds \frac{\mathrm{Disc}f(s)}{s-z}
=\frac{1}{\pi} \int_{4 M^2}^\infty ds \frac{\mathrm{Im}f(s)}{s-z} \; ,
\ee
where in the last step we have used the relation $\mathrm{Disc} f(s)=2i
\mathrm{Im} f(s)$ which holds in the case of the two point function we are
considering. Before applying this formula to the self-energy
$\tilde\Pi_r(s)$, we first need to impose the two renormalization
conditions~\eqref{eq:RenC}. We can do this by subtracting from the integral
its value and its derivative at $s=m^2$. We thus obtain, after inserting
the expression of the imaginary part of $\tilde\Pi_r(s)$ at $\cO(g^2)$,
Eq.~\eqref{eq:ImPi}, the following expression: 
\be
\tilde{\Pi}_r(p^2)=-\frac{\alpha_g}{4} \frac{(p^2-m^2)^2}{\pi} \int_{4M^2}^\infty ds
\frac{\sigma(s)}{(s-m^2)^2(s-p^2)} \; .
\ee
It is not immediate to see that this representation coincides with the one
in Eq.~\eqref{eq:Pirexpl}, but it is easy to realize that the latter is
analytic in the whole complex plane (for complex values of $p^2$ the
integral can be evaluated for $\varepsilon=0$) other than for the cut
starting at $4M^2$. Since the two representations have the same analytic
properties as well as the same discontinuity, they must be identical. A
numerical check can be easily done with the help of a computer-algebra
program and should convince also the skeptics.

In summary, we have shown that unitarity\footnote{Technically we have used
  the completeness of the set of asymptotic states in the derivation above,
  but this property also implies the unitarity of the $S$-matrix, which is
  the word I am using here for simplicity.} and
analyticity provide an alternative derivation of the expression of the
self-energy at $\cO(g^2)$ to the one obtained with a standard loop
calculation. On the other hand, the derivation based on analyticity and
unitarity does not only work in conjunction with perturbation theory but is
more general than that. Indeed we could imagine to add to the Lagrangian in
Eq.~\eqref{eq:LagrY} a self-interaction for the $\phi$ field, for example a
$\lambda | \phi|^4 $ term, with the coupling constant $\lambda \sim
\cO(1)$. Taking this at face value and ignoring issues related to the
proper definition of this as a QFT, we can say that application of a
perturbative expansion in $\lambda$ would not make sense in this case. In
particular, we would not be able to calculate the mass of the meson created
by the $\phi$ field in perturbation theory, nor the rest of the
spectrum. We could still apply perturbation theory in the coupling constant
$g$, however, provided that this stays small. For the self-energy, \eg, it
still makes sense to evaluate it at $\cO(g^2)$. This could get contribution
from multiple intermediate states, $2 \phi$, $4 \phi$ or whatever states
appear in the spectrum: at order $g^2$ the Lagrangian permits a coupling
between the $\varphi$ and two $\phi$ fields, but these can then create any
state thanks to nonperturbative effects in $\lambda$. At this point we do
not have any algebraic principle to establish a hierarchy among these
states, but we can still observe that their thresholds are different and
that each of them can be evaluated individually and gives a positive
contribution. For example, the $2 \phi$-state contributes thanks to the
matrix element in \eqref{eq:2phivac}, which now will have to include a form
factor even at $\cO(g)$: \be \langle 2 \phi| \varphi(0)| 0 \rangle =
\frac{g}{(p^2-m^2)} F_{\varphi \phi \phi}(p^2) \; ,
\label{eq:FFphi}
\ee
a non-perturbative, unknown function of $\lambda$ and $p^2$. The rest of
the argument leading to the dispersive expression of the self-energy
remains the same---the only change is that wherever we had $g$ we have to
replace it with $g F_{\varphi \phi \phi}(p^2)$, which leads to
\be
\tilde{\Pi}^{2 \phi}_r(p^2)=-\frac{\alpha_g}{4} \frac{(p^2-m^2)^2}{ \pi} \int_{4M^2}^\infty ds
\frac{\sigma(s) |F_{\varphi \phi \phi}(s)|^2 }{(s-m^2)^2(s-p^2)} \; .
\label{eq:Piphiphi}
\ee

One could at this point make this simple model more complicated and closer
to the real world, by introducing other kinds of (heavier) particles, with
different spins and quantum numbers, which all couple to the scalar
$\varphi$. For each pair of a new kind of particle one would then get a
contribution completely analogous to that due to the $2 \phi$ state shown
in~\eqref{eq:Piphiphi}. But instead of dwelling further with this toy
model, let us move on and consider the real world.

\subsection{Two-point function of the electromagnetic current in
QCD} 

The toy model above and all the discussion of the two-point function of the
$\varphi$ field can be taken over to the case of the standard model and the
two-point function of the electromagnetic current, and in particular the
hadronic contribution to it. In this case we are also performing a
perturbative expansion in the electromagnetic coupling constant ($\alpha$
instead of $\alpha_g$), but treating the strong interaction
non-perturbatively, as we did with the interaction mediated by $\lambda$
above. The only differences to be accounted for concern the vector nature
of the electromagnetic field, instead of the scalar $\varphi$, and the fact
that the former is massless, $m \to 0$, which is related to
gauge-invariance. The self-energy at $\cO(e^2)$ in this case is given by
\begin{align}
i e^2 \int d^4x e^{iqx} \langle 0 | T j_\mu(x) j_\nu(0) | 0 \rangle & =
\Pi_{\mu \nu}(q^2) \nonumber \\
&=(q^2 g_{\mu \nu}-q_\mu q_\nu) \Pi(q^2) \; ,
\end{align}
and satisfies the renormalization condition $\Pi(0)=0$. Its dispersive
representation, which accounts for the analyticity properties, reads
\be
\Pi(s)=\frac{s}{\pi} \int_{4 M_\pi^2}^\infty ds' \frac{\mathrm{Im}
\Pi(s')}{s'(s'-s)} \; ,
\label{eq:Pi-disp}
\ee
and its imaginary part is fixed by unitarity as
\begin{align}
\mathrm{Im} \Pi(s)&=\sum_n g^{\mu \nu} \langle 0|j_\mu(0)|n \rangle \langle
n|j_\nu(0)|0 \rangle   \nonumber \\
&= \frac{s}{4 \pi \alpha} \sum_n \sigma(e^+e^- \to n) = \frac{s}{4 \pi
\alpha} \sigma(e^+e^- \to \mbox{hadrons}) = \frac{\alpha}{3}R(s) \; ,
\end{align}
where $R(s)=\frac{\sigma(e^+e^- \to \mbox{hadrons})}{\sigma(e^+e^- \to
\mu^+\mu^-)}$, is the well known $R$-ratio, usually defined with the cross
section in the denominator evaluated for massless leptons. 

The hadronic contribution to the photon self-energy is traditionally called
hadronic vacuum polarization (HVP) and plays an important role in two
phenomenologically highly relevant applications:
\begin{enumerate}
\item the running of $\alpha_\mathrm{em}(q^2)$;
\item the anomalous magnetic moment of the muon $(g-2)_\mu$.
\end{enumerate}
The running of $\alpha_\mathrm{em}$ is directly given by $\Pi(s)$ (modulo
simple factors):
\be
\Delta \alpha_\mathrm{em,had}^{(5)}(M_Z^2)=\frac{\alpha M_Z^2}{3
\pi} \Pint_{4 M_\pi^2}^\infty ds \frac{R(s)}{s(M_Z^2-s)} \; ,
\label{eq:alpharun}
\ee
where the 5 in superscript refers to the number of quarks which play a role
in the running (with no constraints in the $R$-ratio all quarks other than
the top contribute), and the bar on the integral sign stands for principal
value integral. 

For precision physics at the electroweak scale this contribution is very
important, because it is the one limiting the precision in the running of
$\alpha$, which allows one to connect precision measurements at $M_Z$ to
precision measurements of $\alpha$ at $q^2=0$. I won't delve into this
subject here, but just want to establish the connection back to a pure
perturbative calculation. Indeed Eq.~\eqref{eq:alpharun} provides an
explicit representation of the $\cO(\alpha)$ contribution of {\em leptons}
to the running of $\alpha$ provided one sets $R(s)=1$ and $4 M_\pi^2 \to
4 m_\ell^2$. In this case the integral can be calculated explicitly:
\be
\frac{\alpha M_Z^2}{3 \pi} \Pint_{4 m_\ell^2}^\infty ds
\frac{1}{s(M_Z^2-s)} = \frac{\alpha}{3 \pi} \ln \left(\frac{M_Z^2}{4
m_\ell^2} \right). 
\ee
This is the kind of large log which needs to be resummed by RGE. As long as
one keeps $R(s)$ in the integrand, the integral cannot be calculated
exactly, and the RGE cannot be solved either. Moreover, if one tries to
approximate the integral with a logarithm of $M_Z$, the effective hadronic
scale which will give the best approximation is rather $M_\rho$ than
$M_\pi$, which means that the log is much smaller than the one for leptons,
which makes the need of a resummation less urgent.

\begin{figure}[t]
\centering
\includegraphics[width=5cm,trim= 8.3cm 0 0 1.7cm,clip]{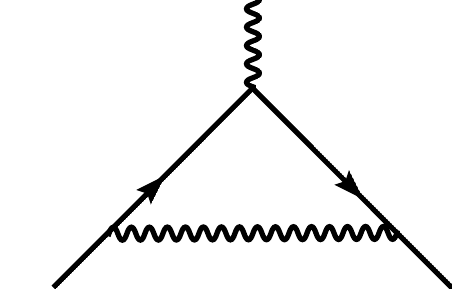}
\hskip 1 cm
\includegraphics[width=4.5cm,trim= 0 0 0 2.5cm, clip]{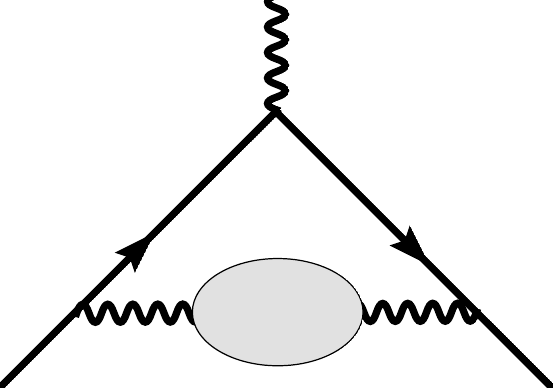}
\caption{Diagrams representing the Schwinger term $\alpha/2 \pi$ (left), the
  leading contribution to the muon $g-2$ and the hadronic vacuum
  polarization (right), the leading, $\cO(\alpha^2)$ hadronic
  contribution. \label{fig:hvp}}  
\end{figure}
The leading contribution to $a_\mu=(g-2)_\mu/2$ is given by the vertex
correction shown on the left-hand side of Fig.~\ref{fig:hvp}. It is finite
and can be calculated analytically, as first done by
Schwinger~\cite{Schwinger:1948iu} almost eighty years ago, with the result:
\be
a_\mu= \frac{\alpha}{2 \pi} + \cO(\alpha^2)=0.00116 \ldots \; .
\ee
Among the $\cO(\alpha^2)$ corrections there is also the leading hadronic,
HVP contribution,  denoted by $a_\mu^\mathrm{HVP}$ and given by the Feynman
diagram on the right-hand side of Fig.~\ref{fig:hvp}. The diagram shows
that in order to calculate the loop integral, one needs to integrate over
$\Pi(\ell^2)$, with $\ell$ the loop four-momentum. Without knowing the
explicit form of the vacuum-polarization function this seems like an
impossible task. On the other hand, if we adopt the
representation~\eqref{eq:Pi-disp}, we see that the only dependence on 
$\ell^2$ appears in the ratio $\ell^2/(s'-\ell^2)$, so that, if we
interchange the loop and the dispersive integral, the former one becomes a
normal, perturbative one-loop integral, similar to the vertex-correction
calculated by Schwinger.
If we include the HVP in the photon propagator, the one-loop integral
increases in complexity going from three to four propagators, but is still 
analytically calculable. The exercise was carried out by Bouchiat and
Michel in 1961~\cite{Bouchiat:1961lbg} and leads to the well-known formula
\be a_\mu^\mathrm{HVP}=\left(\frac{\alpha m_\mu}{3 \pi}\right)^2 \int_{4
  M_\pi^2}^\infty ds \frac{\hat{K}(s)}{s^2} R(s) \; , \ee where
$\hat{K}(s)$ is a smooth function which starts at zero for $s=0$ and grows
monotonically to one for $s \to \infty$, reaching a value of around 0.6
already at $4 M_\pi^2$. Clearly, the $s^{-2}$ factor gives the dominant
energy dependence in the integrand and is responsible for the 
fact that the low-energy region provides the most important contribution. This
observation can be used to make a rough estimate of the relative importance
of the different hadronic contributions. Indeed the imaginary part of
$\Pi(s)$, much like the cross section to which it is proportional, can be
written as a sum of contributions of individual channels: 
\be 
\mathrm{Im} \Pi(s)= \sum_h \mathrm{Im} \Pi_h(s) \; \; \leftrightarrow \; \;
\sigma(e^+e^- \to \mathrm{hadrons})=\sum_h \sigma(e^+e^- \to h) \; .  
\ee
Since each intermediate hadronic state $h$ starts at a different threshold,
there is a suppression in its contribution to $a_\mu$ due mainly to the
factor $1/s^2$: the relative contribution of two different intermediate
states $h_1$ and $h_2$ will scale, in first approximation with
$(s_{h_1}/s_{h_2})^2$ (the ratio of the corresponding energy thresholds
squared) if we ignore the relative size and the energy dependence of
$R_{h_i}(s)$: 
\be 
a_\mu^{\mathrm{HVP},h}=\left(\frac{\alpha m_\mu}{3
    \pi}\right)^2 \int_{s_h}^\infty ds \frac{\hat{K}(s)}{s^2} R_h(s) \; .
\ee 
Indeed the ratio of the thresholds provides a good order-of-magnitude
estimate of the relative size of the contribution of two pions (about $500
\cdot 10^{-10}$) to that of two kaons (about $40 \cdot 10^{-10}$).

\section{Three-point function: form factors and the Omn\`es representation}
In the previous section we have seen that if we want to treat dispersively
the two-point function of an operator $\cO(x)$, we need as input the matrix
elements of this operator between the vacuum and any asymptotic state $| n
\rangle$: $\langle n | \cO(0) | 0 \rangle $. The simplest possible
asymptotic state contributing to the cut is a two-particle state and for
applications like the muon $g-2$, the lower the threshold of the state the
more important its contribution. In QCD and for the muon $g-2$ the most
important matrix element is the one of the electromagnetic current between
the vacuum and the two-pion state, which can be expressed in terms of the
pion vector form factor $F_\pi^V(s)$:
\be
\langle \pi^\pm(p) \pi^\mp(p') \, \mathrm{out} | j^\mu_\mathrm{em}(0) | 0
\rangle=  (p-p')^\mu F_\pi^V(s)\;,  \qquad s=(p+p')^2 \;,
\ee
where
\be
j^\mu_\mathrm{em}(x)= \bar q(x) Q \gamma^\mu q(x) \, , \; \; q=(u,d,s)^T \,
, \; \; Q=\frac{1}{3}\mathrm{diag}(2,-1,-1)\;.
\ee
The modulus of the vector form factor can be measured in $e^+ e^-$
collisions:
\be
\sigma(e^+e^- \to \pi^+ \pi^-)=\frac{\pi \alpha^2}{3 s} \sigma_\pi^3(s)
|F_\pi^V(s)|^2 \; ,
\ee
where $\sigma_\pi(s)\equiv \sqrt{1-4 M_\pi^2/s}$ is the phase-space factor, but
unitarity and analyticity impose constraints on this matrix element 
too, as we are going to discuss in the rest of this section. 

To simplify the derivation of these constraints, I will now concentrate on
the scalar form factor of the pion, which is the matrix element of the
$SU(2)$ scalar and isoscalar operator:
\[
S(x):= \hat{m}\left( \bar{u}(x)u(x)+\bar{d}(x)d(x) \right)\, , \qquad
\hat{m}=\frac{m_u+m_d}{2}  \; ,
\]
namely ($i,j$ are isospin indices of the pion triplet):
\be
\langle \pi^i(p) \pi^j(p') \,\mathrm{out} | S(0) | 0 \rangle=  \delta^{ij}
F_\pi^S(s)\;,  \qquad 
s=(p+p')^2 \;. 
\ee
\vskip 0.3cm
Time-reversal invariance implies that the form factor can equivalently be
defined in terms of the time-reversed process, with an in-state of two
pions being annihilated by the scalar quark bilinear:
\be
\delta^{ij} F_\pi^S(s) = \langle 0 | S(0) | \pi^i(p) \pi^j(p') \,\mathrm{in}
\rangle=\langle \pi^i(p) \pi^j(p') \,\mathrm{in} | S(0) | 0
\rangle^* \; .
\ee
This allows us to express the imaginary part of the form factor as:
\be
\mathrm{Im} F_\pi^S(s)=\frac{1}{2i} \left\{\langle \left(\pi
    \pi\right)_{I=0} \,\mathrm{out} | S(0) | 0 \rangle  
-\langle \left(\pi\pi\right)_{I=0}\,\mathrm{in} | S(0) | 0 \rangle
\right\} 
\ee
after having eliminated the Kronecker-$\delta$ with a projection on the
$2\pi$ $I=0$ state. We can now insert a complete
set of out states between the scalar operator and the two-pion state, which
gives: 
\begin{align}
\mathrm{Im} F_\pi^S(s)&=\frac{1}{2i} \sum_n \left\{\langle
  \left(\pi\pi\right)_{I=0}\,\mathrm{out} |n \,\mathrm{out} \rangle 
-\langle \left(\pi\pi\right)_{I=0}\,\mathrm{in} |n\,\mathrm{out}
\rangle  \right\} \langle n\,\mathrm{out} | S(0) | 0 \rangle \nonumber \\
&=\frac{1}{2i} \sum_n \left\{ \delta_{\left(\pi\pi\right)_{I=0},
    n}-S_{\left(\pi\pi\right)_{I=0} \to n}^*  \right\} \langle
n \,\mathrm{out} | S(0) | 0 \rangle  \nonumber \\
&=\frac{1}{2} \sum_n \left\{ (2 \pi)^4 \delta^4(p+p'-p_n)
  T_{\left(\pi\pi\right)_{I=0} \to n}^*  \right\} \langle 
n \,\mathrm{out} | S(0) | 0 \rangle
\; ,
\end{align}
where in the last step we have made explicit that the $\delta$-function
inside the braces in the second line removes the diagonal part of the
$S$-matrix, leaving nothing but the $T$-matrix, still with its accompanying
$\delta$-function for momentum conservation. If we now concentrate only on
the two-pion contribution to the imaginary part we get:
\be
\mathrm{Im}_{2 \pi} F_\pi^S(s) = \sigma_\pi(s)  t_0^{0\, *}(s) F_\pi^S(s)
\;, 
\ee
where $t^I_\ell(s)$ is the $\pi \pi$ scattering amplitude of angular 
momentum $\ell$ and isospin $I$. The phase-space factor $\sigma_\pi(s)$ is
the result of the integral over all possible momenta of two-pion states
(implicit in the sum over states $n$). Contributions of other intermediate
states $| n \rangle$ to the imaginary part are proportional to the product
between the matrix element $\langle n| S(0) |0 \rangle$ and the scattering
amplitude $\langle 2 \pi \, \mathrm{in}| n \, \mathrm{out} \rangle$, but will
not be discussed here in any detail.

The derivation of the unitarity relation for the vector form factor is
completely analogous and the final expression mirrors the one we just
derived for the scalar form factor:
\be
\mathrm{Im}_{2 \pi} F_\pi^V(s) = \sigma_\pi(s)  t_1^{1\, *}(s) F_\pi^V(s)
\ee
again for the contribution of the two-pion intermediate state.

The question we are going to address is whether by knowing the imaginary
part we can reconstruct the whole form factor. To answer this we have to
ask what are the analyticity properties of the form factor. These are in
fact identical to those of the self-energy: the form factor is an
analytic function in the whole complex plane, with the exception of a cut
on the positive real axis, for $s \ge 4 M_\pi^2$. Below this threshold the
form factor is real on the real axis and it therefore satisfies the Schwarz
reflection principle:
\begin{align}
F_\pi^V(s+i \varepsilon)&=F_\pi^V(s-i\varepsilon)^* \; , \mbox{which
implies} \nonumber \\ 
\lim_{\varepsilon \to 0} F_\pi^V(s+i
\varepsilon)& =:|F_\pi^V(s)| e^{i \delta(s)} \; \; \Rightarrow \; \;
\lim_{\varepsilon \to 0} F_\pi^V(s-i \varepsilon)=|F_\pi^V(s)| e^{-i
\delta(s)} \; ,
\end{align}
{\em i.e.} that the discontinuity can be described as a sign flip of the phase
of the form factor.

We can reformulate the question of the constraints imposed by analyticity
on the form factor as the following mathematical problem: if one knows the
phase of the form factor and its discontinuity is given by a sign flip of
the phase, can one reconstruct the whole form factor, {\em i.e.} express
its modulus through its phase? The solution has been provided by
Omn\`es~\cite{Omnes:1958hv} almost seventy years ago and reads as follows
\be
F(s)=P(s)\Omega(s)\; , \qquad \Omega(s) = \exp\left[ \frac{s}{\pi} \int_{4
M_\pi^2}^\infty ds' \frac{\delta(s')}{s'(s'-s)} \right] \; ,
\ee
where $P(s)$ is a generic polynomial and $\Omega(s)$ is called the Omn\`es
function. This solution further assumes that the form factor does not grow
faster than as a power for $s \to \infty$. This physically reasonable
assumption has later been proved in perturbative
QCD by Brodsky and Lepage~\cite{Lepage:1979zb,Lepage:1980fj}. 
Clearly, the analytic properties remain unchanged no matter what form the
polynomial takes. The polynomial, on the other hand, can be constrained
either by knowing the power of $s$ with which the form factor behaves
asymptotically, or by its zeros, in case these were known. Indeed, if the
phase of the form factor tends to a multiple of $\pi$, $n \pi$, for $s \to
\infty$, $\Omega(s) \sim s^n$, so that, if I know that $F(s) \sim s^m$ in
the same limit, then the polynomial has to be of degree $m-n$. On the other
hand, since the Omn\`es factor has no zeros by construction, if one has
independent information on the number and position of the (complex) zeros,
these have to be generated by the polynomial, whose degree is given by the
number of zeros, and the coefficients are determined by the position of the
zeros, other than for its normalization, which remains unconstrained.

In the case of the vector form factor of the pion, we have information on
its normalization: gauge invariance implies $F_\pi^V(0)=1$, and since by
construction $\Omega(0)=1$, we also have that
$P_\pi^V(0)=1$. Brodsky and Lepage~\cite{Lepage:1979zb,Lepage:1980fj} have
shown that QCD imposes constraints on the behaviour of the form factor at
$s=\infty$, but lacking independent information on the behaviour of the
phase of the form factor, we cannot use this to constrain the degree of the
polynomial. We don't have rigorous information on the presence of zeros for
the vector form factor of the pion, but we can follow the plausibility
argument first formulated by H.~Leutwyler~\cite{Leutwyler:2002hm}: we know
that these cannot be present neither at low energy (because \chpt predicts
that there are none), nor at high energy (where it is perturbative QCD
which forbids them). Assuming that these do not appear, for unknown
reasons, in the region of intermediate energies unaccessible neither to
\chpt nor to pQCD, we can conclude that $P_\pi^V(s)=1$. Analyticity and
some rigorous information from QCD, supplemented by a plausibility
argument, imply that 
\be
F_\pi^V(s)= \Omega_\pi^V(s)
\ee
with the Omn\`es function for this quantity determined by a yet unknown
phase of the vector form factor. The next question is what do we know about
this phase, which can be addressed with the help of unitarity. Indeed we
have seen above that the imaginary part of $F_\pi^V(s)$ is given by the
product of the form factor itself, times the (complex conjugate of the)
$\pi \pi$ scattering amplitude $t_1^1(s)=\sin \delta_1^1(s)
e^{i\delta_1^1(s)}$. Since the imaginary part is by definition real (it
comes multiplied by $i$), we conclude that, below the inelastic threshold,
the phase of the vector form factor coincides with the $\pi \pi$ phase
shift with the same quantum numbers. The following remarks make the Omn\`es
representation of the vector form factor of the pion particularly useful in
the low-energy region, up until about 1
GeV~\cite{Leutwyler:2002hm,Colangelo:2018mtw}:  
\begin{enumerate}
\item
the $\delta_1^1(s)$ phase shift is quite well known, thanks to the Roy
equations and their numerical solution, which will be discussed in the
following section, and the measurement of the
$\sigma(e^+e^- \to \pi^+ \pi^-)$ cross section in the $\rho$-region;
\item
the $3 \pi$ inelasticity, and the corresponding contribution to the vector
form factor, is strongly suppressed by isospin: it is an isospin-violating
effect which is non-negligible only in the region where it is enhanced by
the $\omega$ resonance, where it can be decribed in terms of its
parameters; 
\item
the inelasticity due to $4\pi$ and higher states are phenomenologically
constrained by the Eidelman-\L{}ukaszuk (EL) bound~\cite{Eidelman:2003uh}, which
shows that it starts to be visible only above the $\omega \pi$ threshold,
it starts slowly because of phase-space suppression and remains small until
about 1 GeV and even above. 
\end{enumerate}
Exploiting these considerations we can recast the general Omn\`es
representation in the following form which is useful in the region up to
about 1 GeV~\cite{Leutwyler:2002hm,Colangelo:2018mtw}:
\be
F_\pi^V(s)= \Omega_1^1(s) \cdot G_\omega(s) \cdot \Omega_\mathrm{in}(s) \; ,
\label{eq:FV3F}
\ee
which follows from a splitting of the phase into the sum of three different
contributions:
\be
\delta_\pi^V(s)= \delta_1^1(s)+\delta_{3 \pi}(s)+\delta_\mathrm{in}(s) \; .
\ee
$G_\omega(s)$ is the $3 \pi$ contribution which, given the $\omega$
dominance, can be represented quite accurately by a Breit-Wigner of the
form, $G_\omega(s) \sim g_\omega(s)$: 
\be
g_\omega(s) = 1+ \epsilon_\omega \frac{s}{(M_\omega-\frac{i}{2} \Gamma_\omega)^2-s}
\ee
or, to be more rigorous, by a dispersive integral over its imaginary part
\be
G_\omega=1+ \frac{s}{ \pi} \int_{9 M_\pi^2}^\infty ds'
\frac{\mathrm{Im}g_\omega(s')}{s'(s'-s)} \left(\frac{1-\frac{9
M_\pi^2}{s'}}{1-\frac{9 M_\pi^2}{M_\omega^2} } \right)^4 \; ,
\ee
even though, in practice, the difference between the simplified and the
full-fledged variants of this contribution is almost invisible.

The third factor, $\Omega_\mathrm{in}(s)$, is due to $4 \pi$ and higher
intermediate states, and to the corresponding phase, which the EL-bound
constrains to effectively vanish below $s_\mathrm{in}=(M_\omega+M_\pi)^2$. 
Instead of trying to describe this factor through a phase, which first
needs to be parametrized, and then integrated over, to get the
corresponding Omn\`es factor, we have taken the following shortcut. We
adopted a description in terms of a polynomial in a conformal variable $z$
defined as follows:  
\be
z(s)=\frac{\sqrt{s_\mathrm{in}-s_1}-\sqrt{s_\mathrm{in}-s}}{\sqrt{s_\mathrm{in}-s_1}+\sqrt{s_\mathrm{in}-s}
} 
\ee
where $s_1$ is an arbitrary energy-squared parameter, the value of $s$
which is mapped to $z=0$. This conformal variable maps the complex
$s$-plane to the unit circle in $z$, with $s_\mathrm{in}$ mapped to $z=1$,
the upper (lower) rim of the cut to the upper (lower) half circle, and the
upper (lower) half-plane to the upper (lower) half disk. We then write
\be
\Omega_\mathrm{in}(s)=1+ \sum_{k=1}^N c_k (z(s)^k-z(0)^k)
\ee
with the additional constraint
\be
c_1=-\sum_{k=2}^N k c_k
\ee
to ensure a behaviour like $(s_\mathrm{in}-s)^{3/2}$ near the inelastic
threshold: a polynomial of degree $N$ has $N-1$ free parameters. The
advantage of this representation is that it is easier to use in fits, while
it automatically has the correct analytic properties.

\subsection{Fitting vector form factor data}

\begin{figure}[t!]
\centering
\scalebox{1.0}{\input{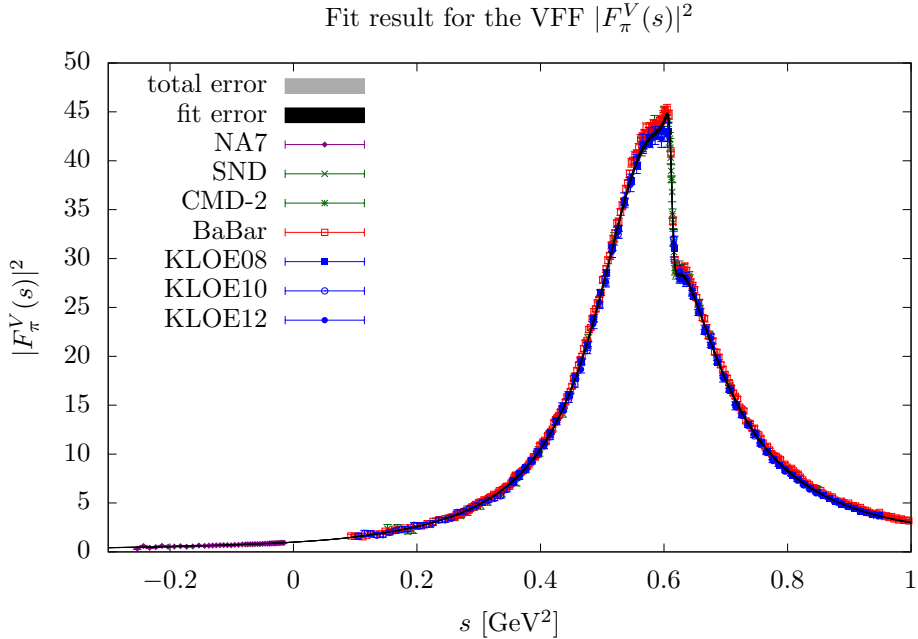}}
\caption{\label{fig:fvfit} Fit to the main sets of timelike and spacelike
  data as of 2018. Figure from Ref.~\cite{Colangelo:2018mtw}.}
\end{figure}
\begin{figure}[t!]
	\centering
	\input{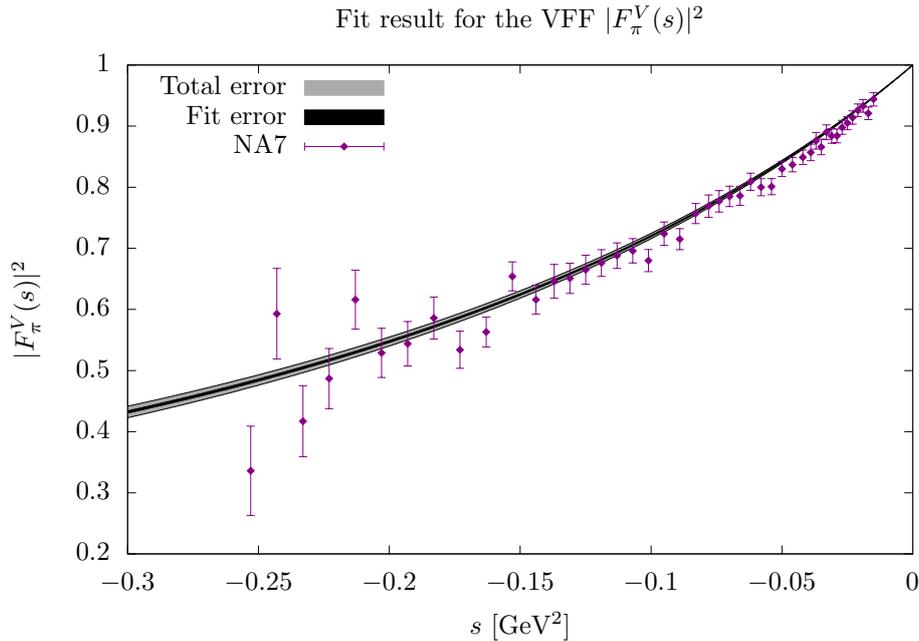}
	\caption{Fit result for the pion VFF in the space-like region,
          together with the NA7 data. Figure from Ref.~\cite{Colangelo:2018mtw}.}
	\label{fig:VFFfitSpacelike}
\end{figure}

In this section we will put this representation to test and see whether it
can describe actual data, essentially by describing the analysis of
Ref.~\cite{Colangelo:2018mtw}. The test is quite non-trivial because
$e^+e^- \to \pi^+\pi^-$ cross section data are very precise, as they have
been collected with the explicit goal of providing an input for the
data-driven HVP contribution to $a_\mu$, for which the precision
requirements are quite high. Figure~\ref{fig:fvfit} displays a common fit
to the main sets of data, both in the timelike as well as in the spacelike
region. The possibility to use the latter data sets is unique to the
dispersive approach, because only when you make explicit use of the
analyticity properties of the vector form factor you can establish a
connection between the two sets of data. Indeed
representation~\eqref{eq:FV3F} can be used without problems in the whole
complex plane, including of course, negative real values of its argument.
Spacelike data have been obtained at CERN by the NA7
Collaboration~\cite{NA7:1986vav} by scattering pions on electrons (of a
liquid hydrogen target), and are better seen in
Fig.~\ref{fig:VFFfitSpacelike} which shows only the spacelike region. While
they don't have a significant impact on the fit, other than a small
reduction of the final uncertainties, it is quite nice to see that the two
sets of data can be linked by analyticity and are mutually consistent.

\begin{figure}[t]
\includegraphics[width=6.5cm]{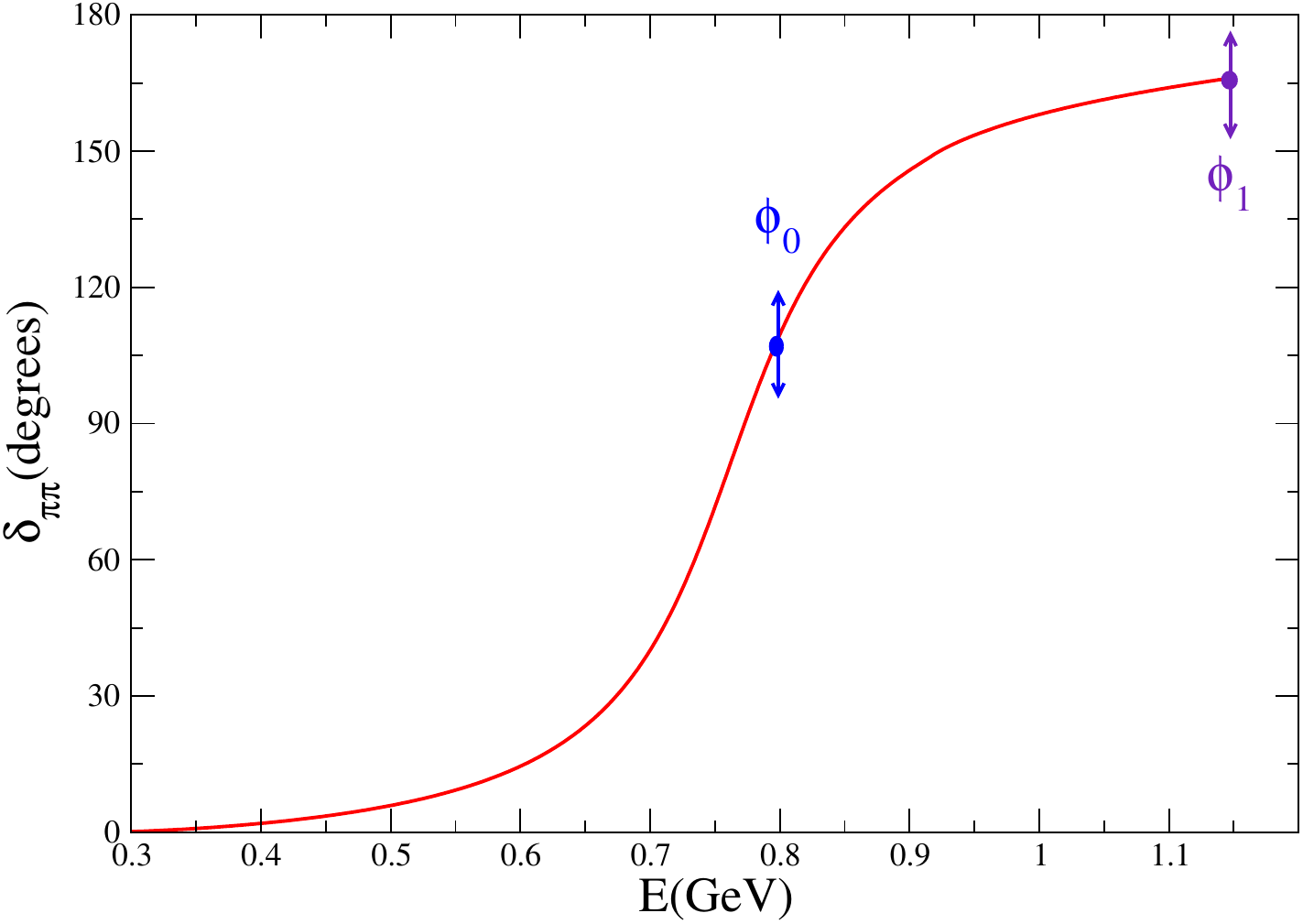}
\includegraphics[width=6.5cm]{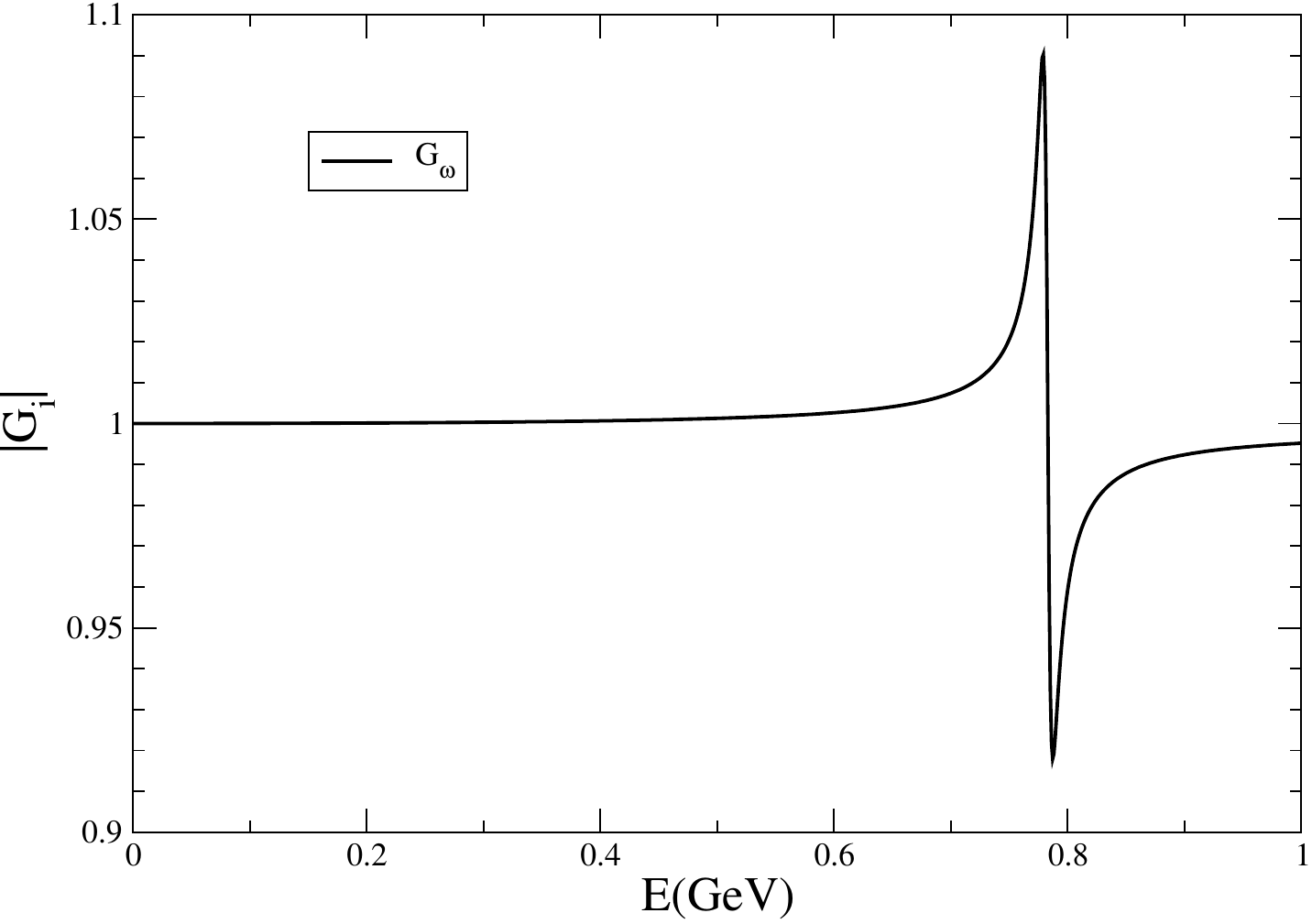}
\caption{\label{fig:d11G2} The left panel shows the two free parameters in the
  elastic part of the Omn\`es function: the value of the $I=1$, $\ell=1$
  phase-shift at two energy points. On the right panel the function
  $|G_\omega(s)|$ is shown: the two parameters $\epsilon_\omega$ and
  $M_\omega$ fix this contribution.}
\end{figure}

\begin{figure}[t]
	\centering
	\scalebox{0.95}{
	\input{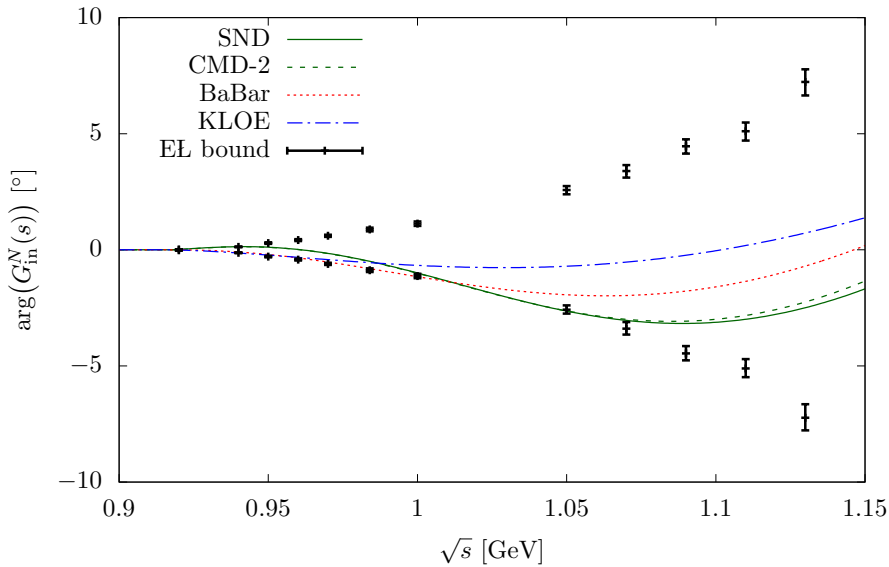}}%
	\caption{Phase of the inelastic contribution $G_\mathrm{in}^N(s)$
          for $N-1=4$ free parameters, shown together with the upper and
          lower \EL{} bounds indicated by the two mirror sets of data
          points. Figure from Ref.~\cite{Colangelo:2018mtw}.} 
	\label{fig:phasediff}
\end{figure}

From the plots it is clear that the dispersive representation can describe
the data well, but for a more quantitative assessment of the quality of the
fit a table of numbers is better than a figure densely packed with data.
In Table~\ref{tab:FinalFitsSingleExperiments} details of the fits to the
individual $e^+e^-$ experiments are shown. These are final fits performed
with four free parameters in the conformal polynomial, and allowing for a
small rescaling of the measured center-of-mass energy, since this is also
subject to a systematic uncertainty. The quality of the fits is shown by
the $\chi^2/$dof reported in the table and is quite acceptable for all data
sets.\footnote{In one of the three datasets of KLOE, the one published in
  2008, there were two datapoints which contributed significantly to the
  $\chi^2$. In KLOE$''$ these two datapoints have been removed, but the fit
  results are barely affected. For details on this and many other aspects,
  see Ref.~\cite{Colangelo:2018mtw}.}
In Table~\ref{tab:FinalFitsSingleExperiments} the essential parameters of
the fits are shown, as illustrated graphically in Fig.~\ref{fig:d11G2}: the
mass of the omega resonance $M_\omega$ and its coupling to the rho, which
determine the factor $G_\omega(s)$, and the value of the $\delta^1_1$
phase-shift at the two energy values $\sqrt{s_0}=0.8$ GeV and
$\sqrt{s_1}=1.15$ GeV, which determine the elastic part of the Omn\`es
function $\Omega_1^1(s)$. Of course also the coefficents of the conformal
polynomial play a role in the fits, but the Eidelman-\L{}ukaszuk bound
constrains this contribution quite tightly, as shown in
Fig.~\ref{fig:phasediff}. 
\begin{table}[t]
	\centering
	\small
	\begin{tabular}{l r r r r r }
	\toprule
		& $\chi^2/$dof$\qquad$ & $M_\omega$ [MeV] 	&	$\delta_1^1(s_0)$ [$^\circ$] & $\delta_1^1(s_1)$ [$^\circ$] & $10^3 \times \epsilon_\omega$  \\
	\midrule
	SND		&	$51.9 / 37 = 1.40$	&	$781.49(32)(2)$	&	$110.5(5)(8)$	&	$165.7(0.3)(2.4)$	&	$2.03(5)(2)$	\\
	CMD-2	        &	$87.4 / 74 = 1.18$	&	$781.98(29)(1)$	&	$110.5(5)(8)$	&	$166.4(0.4)(2.4)$	&	$1.88(6)(2)$	\\
	BaBar	        &	$299.1 / 262 = 1.14$	&	$781.86(14)(1)$	&	$110.4(3)(7)$	&	$165.7(0.2)(2.5)$	&	$2.04(3)(2)$	\\
	KLOE$''$	&	$222.5 / 185 = 1.20$	&	$781.81(16)(3)$	&	$110.3(2)(6)$	&	$165.6(0.1)(2.4)$	&	$1.98(4)(1)$	\\
	\bottomrule
	\end{tabular}
	\caption{Final fits to single $e^+e^-$ experiments with $N-1=4$
          free parameters in the conformal polynomial. The first error is
          the fit uncertainty, inflated by $\sqrt{\chi^2/\mathrm{dof}}$,
          the second error is the combination of all systematic
          uncertainties.} 
	\label{tab:FinalFitsSingleExperiments}
\end{table}

Comparing the values of the inelastic phase with those of the elastic one,
one can see that the former represents a correction of a few percent.
Beyond the size, the energy dependence of this correction plays also an
important role, but in view of the high inelastic threshold, there are only
limited possibilities of generating anything other than a very smooth
energy dependence in the rho-peak region, which is the most important one.
This explains why I have defined the parameters determining $\Omega_1^1(s)$
and $G_\omega(s)$ as the essential fit parameters. Details about the values
of the coefficients of the conformal polynomial obtained in the fit can be
found in Ref.~\cite{Colangelo:2018mtw}.
\begin{figure}[t]
	\centering
        \scalebox{1.0}{
	\input{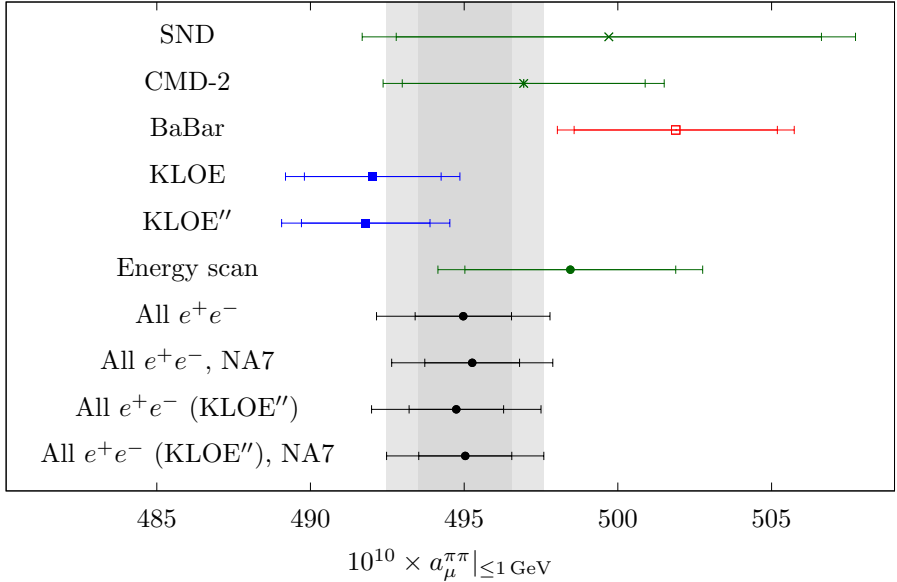}}
	\caption{Results for $a_\mu^{\pi\pi}$ in the energy range
          $\le1\GeV$. The smaller error bars are the fit uncertainties,
          inflated by $\sqrt{\chi^2/\mathrm{dof}}$, the larger error bars
          are the total uncertainties. The gray bands indicate the final
          result. Figure from Ref.~\cite{Colangelo:2018mtw}.} 
	\label{fig:AmuResults}
\end{figure}
\begin{figure}[h]
	\centering
	\includegraphics[width=14cm]{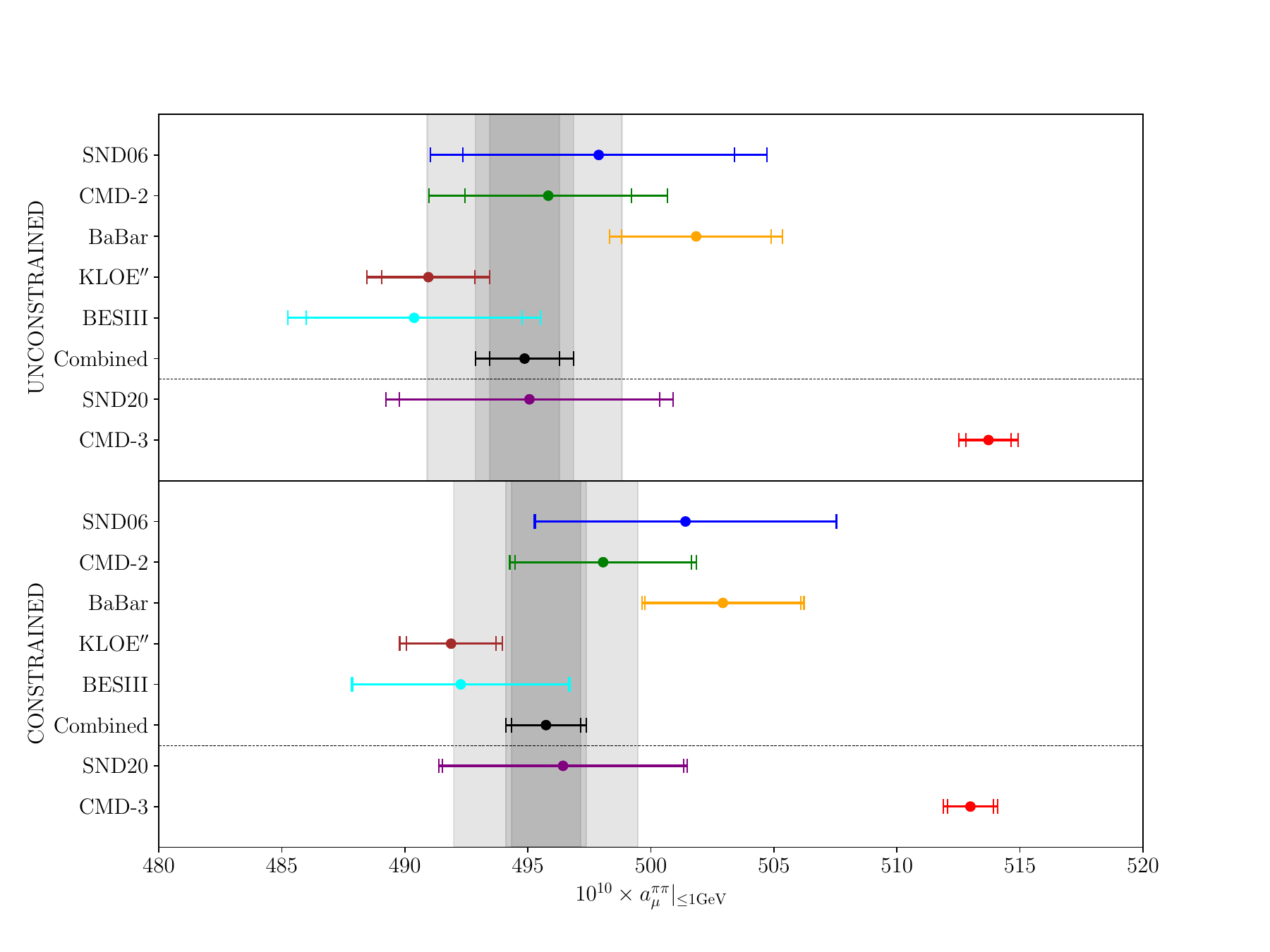}
	\caption{Results for $a_\mu^{\pi\pi}$ in the energy range
          $\le1\GeV$. The smaller error bars show the fit uncertainties,
          inflated by $\sqrt{\chi^2/\mathrm{dof}}$, the larger error bars
          are the total uncertainties, including the systematic ones. The
          gray bands correspond to the combined fit to all $e^+e^-$ data
          sets (apart from SND20 and CMD-3) nad the NA7 set, with the
          largest band including the additional systematic effect due to
          the BaBar-KLOE tension. The upper (lower) panel shows fits
          in which the presence of zeros is not (is) excluded. Figure from
          Ref.~\cite{Leplumey:2025kvv}.} 
	\label{fig:AmuResults-LS}
\end{figure}

As mentioned above, the fit to $e^+e^- \to \pi^+ \pi^-$ data is very
important in view of the determination of the HVP contribution to $a_\mu$.
Having fully determined the pion vector form factor from these data, it is
a trivial exercise to integrate its modulus squared to obtain the
contribution to $a_\mu$, up to 1 GeV. The results, obtained in 2018, are
shown in Fig.~\ref{fig:AmuResults}.  Much has happened in this field in the
last seven years, both on the experimental as well as on the theoretical
side, and an update of Fig.~\ref{fig:AmuResults} is necessary if one wants
to have a better picture of the current situation. In particular because of
the impact of a new measurement of the $e^+e^- \to \pi^+ \pi^-$ cross
section performed by the CMD-3 collaboration in
2023~\cite{CMD-3:2023alj,CMD-3:2023rfe}, which turned out to be higher than
all the other earlier measurements, making all previous tensions visible in
Fig.~\ref{fig:AmuResults} among the different $e^+e^-$ experiments pale in
comparison .

The dispersive analysis of the pion vector form factor and a corresponding
fit to old and new datasets has been recently updated by Leplumey and
Stoffer~\cite{Leplumey:2025kvv}, who in addition have addressed the
question of the presence of zeros in the form factor. As we have discussed
above, the knowledge about the phase of the form factor along the cut fixes
it up to the presence of a polynomial, which in turn can be completely
characterized in terms of its zeros. Even if one knows the phase of the
form factor the presence and the number of zeros may affect the shape of
the form factor. The parametrization discussed above does not completely
exclude the presence of zeros, but the only factor which may have any is
the conformal polynomial. Theoretical arguments against the possible
presence of zeros have been sketched above, following
Leutwyler~\cite{Leutwyler:2002hm}, but the fits shown in
Ref.~\cite{Colangelo:2018mtw} have not excluded them. In fact, an a
posteriori analysis has shown that fits with conformal polynomials of
degree higher than a few units did display zeros. The analysis performed by
Leplumey and Stoffer shows what is the impact of excluding zeros in the
conformal polynomial, and this can be seen in
Fig.~\ref{fig:AmuResults-LS}. Excluding zeros shifts the central values of
$a_\mu^{\pi\pi}$ a little and reduces somewhat the uncertainties.

Both the old as well as the most recent analyses of $e^+e^-$ data in view
of a determination of the $\pi \pi$ HVP contribution to $a_\mu$ show the
advantages of a dispersive approach. If one relies only on data to
calculate the integral, one can only calculate it in the region where data
are present. Data below 0.6 GeV are scarce and have significantly larger
error bars, which makes an evaluation of the contribution from threshold up
to this energy quite difficult, and plagued by large uncertainties. Indeed
in the CMD-3 papers~\cite{CMD-3:2023alj,CMD-3:2023rfe} as well as in
several other ones the comparison between different experiments is usually
done in the range between 0.6 and 0.9 GeV. With the dispersive approach it
is sufficient to have very precise data in a finite region to pin down all
the fit parameters and allow for an extrapolation all the way down to
threshold with very limited additional uncertainties. This also emphasizes
the differences between experiments which have a discrepancy only in a
limited region: according to the dispersive description if there is a
disagreement in a certain region, this has to continue also outside that
region, as a consequence of analyticity. Indeed the discrepancies shown in
Fig.~\ref{fig:AmuResults-LS} are quite a bit larger than those shown by
analyses which rely on data only and do not make use of a dispersive
description of the pion form factor. The interested reader is referred to
the latest White Paper of the Muon $g-2$ Theory
Inititative~\cite{Aliberti:2025beg} for more details and for a comparison
among the results shown here and those obtained with purely data-driven
analyses.

\section{Four-point function: Roy equations for a $\lambda \phi^4$ theory}
In the previous section we have seen that in a dispersive representation of
the vector form factor of the pion the essential parameters (in particular
in the isospin limit) are the values of the phase-shift with the relevant
quantum numbers at two energy points. One of them is fixed at 1.15 GeV
(above this point the phase-shift is determined experimentally), but the
second, lower one can be chosen arbitrarily. The crucial information is
that if the phase is known at these two values of its argument (and above
1.15 GeV), then it is known everywhere. The shape of the phase-shift curve
between threshold and the given value at 1.15 GeV is completely fixed in
terms of its value at a single point in between. This remarkable conclusion
can be reached after considering the analyticity and unitarity constraints
on the $\pi \pi$ scattering amplitude, which lead to the so-called Roy
equations, which were first derived by S.M.~Roy in 1971~\cite{Roy:1971tc}.
The derivation of these equations is somewhat obscured by the presence of
isospin and the corresponding indices which need to be attached to pions.
These make the crossing relations more cumbersome than necessary if one
just wants to understand the essence of the derivation of the Roy
equations. For this reason I will first derive these equations for a theory
with a single, self-interacting, real scalar. To fix the ideas the reader
can imagine that this is a $\lambda \phi^4$ theory.

\begin{figure}[t]
\centering
\includegraphics[width=13cm, trim=0 14cm 0 0,clip]{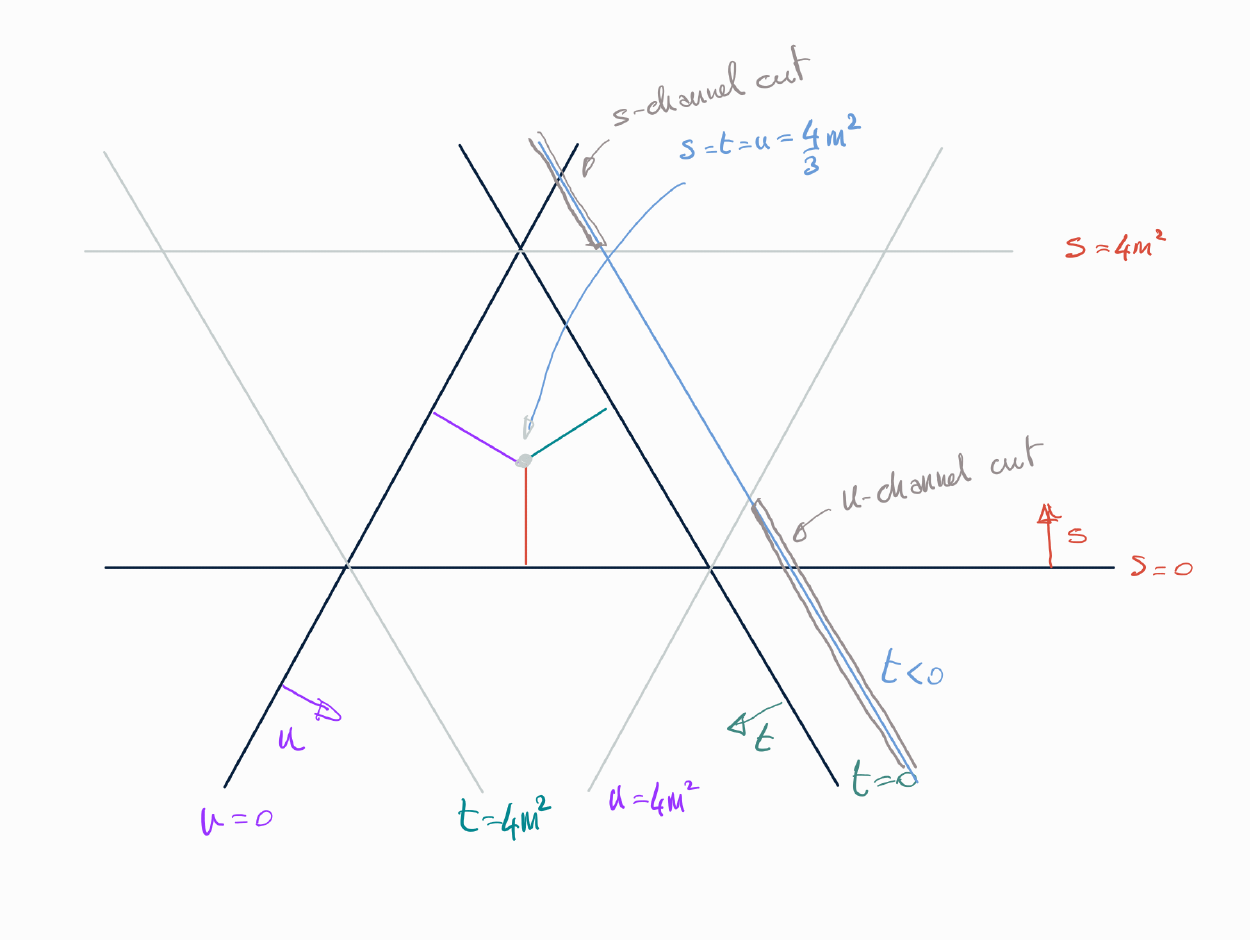}
\caption{Mandelstam plane for elastic scattering of two particles of equal
  mass $m$. The line of constant $t<0$ is the one along which the fixed-$t$
  dispersion relation is derived. \label{fig:Mstam-plane}}
\end{figure}
We consider now the scattering amplitude of this scalar particle (of mass
$m$): 
\be
\langle \phi(p_3) \phi(p_4)\, \mathrm{out} | \phi(p_1) \phi(p_2) \,
\mathrm{in}\rangle
= \delta_{fi}+i (2 \pi)^4 \delta^4(P_f-P_i) A(s,t,u)
\ee
where $s=(p_1+p_2)^2$, $t=(p_1-p_3)^2$ and $u=(p_1-p_4)^2$ are the usual
Mandelstam variables: $s+t+u=4m^2$. Crossing relations imply in this case
that the amplitude is completely crossing symmetric:
\be
A(s,t,u)=A(t,u,s)=A(s,u,t) \; .
\ee
In each channel the amplitude develops an imaginary part in the physical
region, $x\ge 4 m^2$, for $x=s,t,u$. To simplify the analysis it is
convenient to fix one of the three variables and consider the dependence
only on the remaining independent one. To be concrete, we will fix the
variable $t$ at a value between $0$ and $- 4 m^2$. In this case, as we
vary the variable $s$, we move from the region where $s > 4m^2$, and the
imaginary part is generated by rescattering of particles in the $s$
channel, to the region $s <-t$, where $u> 4 m^2$ and the imaginary part is
generated by the rescattering in the $u$ channel. This is illustrated in
Fig.~\ref{fig:Mstam-plane}.

These imaginary parts are cuts, over which we can integrate to reconstruct
the whole amplitude dispersively. Adopting a dispersion relation with a
single subtraction (just to simplify the formulae), we can write:
\be
A(s,t,u)=c(t)+\frac{s}{\pi} \int_{4 m^2}^\infty ds'
\frac{\mathrm{Im}A(s',t,u')}{s'(s'-s)} +\frac{u}{\pi} \int_{4 m^2}^\infty du'
\frac{\mathrm{Im}A(s',t,u')}{u'(u'-u)}
\ee
where the $u$ variable, primed or not, has to be viewed as dependent:
$u^{(\prime)}=4m^2-s^{(\prime)}-t$. Since the amplitude is completely
symmetric in all its variables, I can rewrite the second dispersive
integral as
\be
\frac{u}{\pi} \int_{4 m^2}^\infty ds'
\frac{\mathrm{Im}A(s',t,u')}{s'(s'-u)} \; ,
\ee
which leads to
\be
A(s,t,u)=c(t)+\frac{1}{\pi} \int_{4 m^2}^\infty \frac{ds'}{s'}
\left(\frac{s}{s'-s}+\frac{u}{s'-u} \right) \mathrm{Im}A(s',t,u') 
\label{eq:Afixedt}
\ee
and makes the $s \leftrightarrow u$ crossing symmetry explicit.
Notice, however, that the subtraction constant is in this case a function
of $t$ and that the special treatment of this variable prevents us from
making the full crossing symmetry evident. However, we can move in this
direction if we exploit crossing symmetry to obtain an explicit expression
of $c(t)$. To do so, we consider the amplitude at $s=0$ and $u=u_0:=4m^2-t$:
\be
A(0,t,u_0)=c(t)+\frac{u_0}{\pi} \int_{4 m^2}^\infty
ds'\frac{\mathrm{Im}A(s',t,u')}{s'(s'-u_0)} \; .
\ee
But crossing symmetry implies that $A(0,t,u_0)=A(t,0,u_0)$, and the
latter can be expressed as:
\be
A(t,0,u_0)=c(0)+\frac{t}{\pi} \int_{4 m^2}^\infty ds'
\frac{\mathrm{Im}A(s',0,u_0')}{s'(s'-t)}+\frac{u_0}{\pi} \int_{4 m^2}^\infty ds'
\frac{\mathrm{Im}A(s',0,u_0')}{s'(s'-u_0)}
\ee
which leads to the following expression for $c(t)$:
\begin{align}
c(t)=&c(0)+\frac{t}{\pi} \int_{4 m^2}^\infty ds'
\frac{\mathrm{Im}A(s',0,u_0')}{s'(s'-t)} \nonumber \\
&\quad \;\;\, +\frac{u_0}{\pi} \int_{4 m^2}^\infty ds'
\frac{\mathrm{Im}A(s',0,u_0')-\mathrm{Im}A(s',t,u')}{s'(s'-u_0)} \; .
\end{align}
Inserting this into Eq.~\eqref{eq:Afixedt} we obtain
\begin{align}
A(s,t,u)&=c(0)+\frac{1}{\pi}\int_{4 m^2}^\infty \frac{ds'}{s'}
\left(\frac{s}{s'-s}+\frac{t}{s'-t}+\frac{u}{s'-u} \right)
\mathrm{Im}A(s',t,u') \nonumber \\ 
&\qquad \quad +\frac{t}{\pi} \int_{4 m^2}^\infty ds'
\frac{\mathrm{Im}A(s',0,u_0')-\mathrm{Im}A(s',t,u')}{s'(s'-t)} \nonumber \\
&\qquad \quad +\frac{u_0}{\pi} \int_{4 m^2}^\infty ds'
\frac{\mathrm{Im}A(s',0,u_0')-\mathrm{Im}A(s',t,u')}{s'(s'-u_0)} \; ,
\label{eq:AstuDR}
\end{align}
which has a first line which displays a manifest crossing symmetry, whereas
the rest doesn't---this is clearly only a matter of the crossing symmetry
being manifest or not in this representation, as there is no doubt that
this amplitude satisfies this property.

\subsection{Partial wave projection}
If we are interested in partial waves we can simply apply the standard
projection:
\be
t_\ell(s)=\frac{1}{2} \int_{-1}^1dz P_\ell(z)A(s,t(s,z))
\ee
with $P_\ell(z)$ the Legendre polynomials and $t(s,z)=\frac{1}{2}(4
m^2-s)(1-z)$. In view of crossing symmetry, the projection can also be
obtained by integrating only over half of the domain:
\be
t_\ell(s)=\int_{0}^1dz P_\ell(z)A(s,t(s,z))
\ee
which is convenient because it involves a smaller range of values for $t$.
Applying this projection to Eq.~\eqref{eq:AstuDR} one obtains a dispersion
relation for the partial waves, but on the right-hand side of the equation
one would still have the imaginary part of the full scattering amplitude
rather than its partial waves. To have the partial waves appear also in the
integrand, one has to decompose the imaginary parts of the scattering
amplitudes in the dispersive integrals into partial waves too. This
involves a different projection because the relevant variables are now $s'$
and $t$: 
\be
\mathrm{Im}A(s',t)=\sum_{\ell=0}^\infty (2 \ell + 1)
P_\ell\left(1+\frac{2t}{s'-4m^2} \right) \mathrm{Im} t_\ell(s') \; .
\ee
Inserting this expression in the dispersive integrals and carrying out
the projection onto partial waves for the external kinematic variables one
obtains 
\be
t_\ell(s)=a_0 \delta_{\ell 0}+ \sum_{\ell'=0}^\infty \int_{4 m^2}^\infty
ds' K_{\ell \ell'}(s,s') \mathrm{Im} t_{\ell'}(s') \; ,
\ee
where $a_0$ is the $S$-wave scattering length: $t(4 m^2)=a_0$. The choice
of the subtraction point is arbitrary and has no consequence on the
physics. A change from one subtraction constant to a different one can also
be done a posteriori and the relation between two different subtraction
constants can be expressed as a sum of integrals over kernel functions
multiplying the imaginary parts. With this choice of subtraction point the
kernel functions have to satisfy: $K_{\ell \ell'}(4 m^2,s')=0$. I will not
provide here the explicit form of the kernel functions, but this is a
useful exercise for the interested reader.

\section{Four-point function: Roy equations for pion scattering}
The physical case of $\pi \pi$ scattering has one additional
complication with respect to the single-scalar theory discussed above:
pions build an isospin triplet and, correspondingly, the 
scattering amplitude has isospin indices. If isospin is conserved, as I am
assuming here, a generic pion scattering amplitude can be written as: 
\begin{align}
&\langle \pi^c(p_3) \pi^d(p_4) \, \mathrm{out} | \pi^a(p_1)
\pi^b(p_2) \, \mathrm{in}\rangle=  \hskip 2.5cm [a,b,c,d=1,2,3]\\
&= \delta_{fi}+i (2 \pi)^4 \delta^4(P_f-P_i) \left[\delta^{ab}\delta^{cd}
  A(s,t,u)+\delta^{ac}\delta^{bd} A(t,u,s)+\delta^{ad}\delta^{bc} A(u,s,t)
\right] \;,  \nonumber
\end{align}
with $A(s,t,u)$ the isospin-invariant scattering amplitude, which is
symmetric under the exchange of the last two arguments
$A(s,t,u)=A(s,u,t)$. 
Any scattering amplitude can be expressed in term of the isospin invariant
one. For example, the three amplitudes with a fixed isospin in the $s$
channel can be written as
\begin{align}
T^0(s,t)&=3 A(s,t,u)+A(t,u,s)+A(u,s,t) \nonumber \\
T^1(s,t)&=  A(t,u,s)-A(u,s,t) \label{eq:TIs} \\
T^2(s,t)&=  A(t,u,s)+A(u,s,t) \nonumber \; .
\end{align}
The presence of isospin makes crossing relations more complicated: an
isospin eigenstate of two pions scattering in the $s$ channel will be,
in general, a mixed isospin state in the $t$ channel. Such crossing relations
are linear and can be expressed in terms of crossing matrices according to
the following conventions. First: we build a vector of amplitudes out of
the three isospin scattering amplitudes $T^I(s,t)$ and establish the
convention that the isospin state refers to the channel corresponding to
the first argument of the amplitude. So $T^1(s,t)$ is the amplitude of
isospin 1 in the $s$ channel, but $T^1(t,s)$, the one of isospin 1 in the
$t$ channel.  
\be
\vec{T}(s,t)= \left( \begin{array}{l} T^0(s,t) \\ T^1(s,t) \\ T^2(s,t)
  \end{array} \right) \, , \qquad 
\vec{T}(t,s)= \left( \begin{array}{l} T^0(t,s) \\ T^1(t,s) \\ T^2(t,s)
  \end{array} \right) \,,  \; \; \mbox{etc.}
\ee
With these conventions crossing relations read as follows:
\begin{align}
\vec{T}(s,u)&= C_{tu} \vec{T}(s,t) \nonumber \\
\vec{T}(t,s)&= C_{st} \vec{T}(s,t) \\
\vec{T}(u,t)&= C_{su} \vec{T}(s,t) \nonumber
\end{align}
with the crossing matrices which can be easily calculated from
Eq.~\eqref{eq:TIs} after applying the necessary permutations of
arguments. Their expressions read:
\be
C_{tu}=\left( \begin{array}{rrr} 1&0&0 \\ 0&-1&0\\ 0&0&1  \end{array}
\right) \, , \; \; 
C_{st}=\left( \begin{array}{rrr} 1/3&1&5/3 \\ 1/3&1/2&-5/6\\ 1/3&-1/2&1/6  \end{array}
\right) \, , \; \; 
C_{su}=\left( \begin{array}{rrr} 1/3&-1&5/3 \\-1/3&1/2& 5/6\\ 1/3&1/2&1/6  \end{array}
\right) \, .
\ee
Equipped with this formalism we can now repeat the same steps performed for
the case of the single scalar, after adapting them to the presence of
isospin and the corresponding crossing relations.

In this case the fixed-$t$ dispersion relation with two subtractions reads
as follows:
\begin{align}
\vec{T}(s,t)&=C_{st}\left[\vec{C}(t)+(s-u) \vec{D}(t) \right] \nonumber \\
&+\frac{1}{\pi}
\int_{4 M_\pi^2}^\infty \frac{ds'}{s^{\prime 2}}
\left(\frac{s^2}{s'-s}+\frac{u^2}{s'-u} C_{su} \right) \mathrm{Im}
\vec{T}(s',t) \; .
\label{eq:TfixedtDR}
\end{align}
The detailed derivation of this expression is left to the reader, but the
following two remarks should be helpful in this regard. First: the presence
of the $C_{st}$ matrix in front of the subtraction term can be understood
by considering that the subtraction coefficients $\vec{C}$ and $\vec{D}$
are isospin vectors and depend only on $t$: their isospin therefore refers
to the $t$ channel, but we are considering the amplitude with fixed isospin
in the $s$ channel. This crossing operation is trivially performed with
multiplication by the $C_{st}$ matrix. Moreover, the term linear in $s$
and in $u$ can be decomposed into an even and an odd part under the
exchange of $s$ and $u$. The even part is proportional to $s+u=4
M_\pi^2-t$, and as such is only a function of $t$, which can be fully
reabsorbed in the ``constant'' $\vec{C}(t)$. This explains why the only
linear term is proportional to $s-u$ and why the isospin decomposition of
the subtraction coefficients looks as follows:
\be
\vec{C}(t)=\left( \begin{array}{c} c^0(t)\\ 0 \\ c^2(t) \end{array} \right)
\, , \; \; \; 
\vec{D}(t)=\left( \begin{array}{c} 0\\ d^1(t) \\ 0      \end{array} \right)
\; .
\label{eq:CDt}
\ee
The second observation concerns the presence of the $C_{su}$ crossing
matrix in the second term in the dispersive integral. It should be quite
obvious that this matrix is needed: we are integrating over the cut in the
$u$ channel and the $C_{su}$ matrix is what gives us the amplitudes of
fixed isospin in the $u$ channel starting from the amplitudes of fixed
isospin in the $s$ channel. In this way, we can write the integrand as
being proportional (in a matrix sense) to the imaginary part of the vector
of amplitudes of fixed isospin in the $s$ channel.

In the next step we have to exploit crossing symmetry to reexpress the
subtraction coefficients in terms of true constants and dispersive
integrals. One considers that
\be
\vec{T}(0,t)=C_{st} \vec{T}(t,0)
\ee
and uses Eq.~\eqref{eq:TfixedtDR} to give explicit expressions to both
sides of the equation:
\begin{align}
\vec{T}(0,t)&=C_{st}\left[ \vec{C}(t)-u_0 \vec{D}(t) \right]+\frac{1}{\pi} \int_{4 M_\pi^2}^\infty \frac{ds'}{s^{\prime 2}}
\left(\frac{u_0^2}{s'-u} C_{su} \right) \mathrm{Im} \vec{T}(s',t) \\
\vec{T}(t,0)&=C_{st}\left[\vec{C}(0)+(t-u_0) \vec{D}(0) \right]+\frac{1}{\pi}
\int_{4 M_\pi^2}^\infty \frac{ds'}{s^{\prime 2}}
\left(\frac{t^2}{s'-t}+\frac{u_0^2}{s'-u_0} C_{su} \right) \mathrm{Im}
\vec{T}(s',0) \, .  \nonumber
\end{align}
Exploiting the fact that the two vectors $\vec{C}(t)$ and $\vec{D}(t)$ are
orthogonal, see~\eqref{eq:CDt}, we can now solve the equation for both of
them. The final step (which we didn't do explicitly for the case of the
single scalar) is to reexpress the subtraction constants in terms of the
scattering lengths. After doing this one ends up with:
\begin{align}
\vec{T}(s,t)&= \frac{1}{4 M_\pi^2} \left( s \mathbf{1}+ t C_{st}+ u C_{su}
\right) \vec{T}(4 M_\pi^2,0) \nonumber \\
&+\int_{4 M_\pi^2}^\infty ds' g_2(s,t,s') \mathrm{Im} \vec{T}(s',0) 
+\int_{4 M_\pi^2}^\infty ds' g_3(s,t,s') \mathrm{Im} \vec{T}(s',t)
\end{align}
where
\begin{align}
g_2(s,t,s')&=-\frac{t}{\pi s'(s'-4M_\pi^2)} \left(u C_{st}+s C_{st} C_{tu}
\right) \left( \frac{ \mathbf{1}}{s'-t} +\frac{C_{st}}{s'-u_0} \right)
\nonumber \\
g_3(s,t,s')&=-\frac{su}{\pi s'(s'-u_0)} \left( \frac{ \mathbf{1}}{s'-s}
  +\frac{C_{su}}{s'-u} \right) \; .
\end{align}

We now proceed and apply partial-wave projection both to the isospin
amplitude on the right-hand side of the equation and to the imaginary part
in the integrand. To be able to compare to expressions in the literature
one needs to follow the same conventions, which are given explicitly here: 
\be
T^I(s,t)= 32 \pi \sum_{\ell=0}^\infty (2 \ell +1) P_\ell \left(1+
  \frac{2t}{4 M_\pi^2-s} \right) t^I_\ell(s) \; .
\ee
The partial waves are usually expressed in terms of phase-shift and
inelasticity: 
\be
t^I_\ell(s)= \frac{1}{2i\sigma(s)} \left[\eta^I_\ell(s)e^{2i
    \delta^I_\ell(s)} -1 \right] \; ,
\ee
with $\sigma(s)=\sqrt{1- 4M_\pi^2/s}$. 
The Roy equations for the partial waves read then as follows:
\be
t^I_\ell(s)=k^I_\ell(s)+\sum_{I'=0}^2 \sum_{\ell'=0}^\infty \int_{4
  M_\pi^2}^\infty ds' K^{II'}_{\ell \ell'}(s,s')
\mathrm{Im}t^{I'}_{\ell'}(s') \; , 
\label{eq:Royfull}
\ee
with
\be
k^I_\ell(s)=a_0^I \delta^0_\ell+\frac{s-4M_\pi^2}{4 M_\pi^2}(2 a_0^0-5
a_0^2) \left(\frac{1}{3} \delta^I_0 \delta^0_\ell +\frac{1}{18} \delta^I_1
  \delta^1_\ell -\frac{1}{6} \delta^I_2 \delta^0_\ell \right) \; .
\ee
The kernels in the integrand can be easily obtained from the kernel
functions $g_{2,3}(s,t,s')$ after internal and external partial wave
projections. Their expression reads
\be
K^{II'}_{\ell \ell'}(s,s')=(2 \ell'+1) \int_0^1 dz P_\ell(z) \left[
g_2^{II'}(s,t_z,s')+g_3^{II'}(s,t_z,s')P_{\ell'}\left(1+\frac{2 t_z}{s'-4M_\pi^2} \right)
\right] \; ,
\ee
with $t_z=1/2(4 M_\pi^2-s)(1-z)$, and where we have made explicit the
isospin indices of the kernel functions $g_{2,3}$, as these are then taken
over by the kernels $K^{II'}_{\ell \ell'}$.

Inserting this expression into a computer algebra system makes the
calculation of any kernel completely trivial, but for illustration I will
provide here the explicit expression for one of them, that with
$I=I'=\ell=\ell'=0$:
\begin{align}
K^{II'}_{\ell \ell'}(s,s')&=\frac{\delta^{II'} \delta_{\ell \ell'}}{\pi
  (s'-s)} + \bar{K}^{II'}_{\ell \ell'}(s,s') \nonumber \\
\bar{K}^{00}_{00}(s,s')&=\frac{2}{3 \pi (s-4M_\pi^2)} \ln \left( \frac{s+s'-4
    M_\pi^2}{s'}\right) -\frac{2s+5 s'-16 M_\pi^2 }{3 \pi s' (s'- 4M_\pi^2) }
 \; .
\label{eq:K0000}
\end{align}
From this expression one can see that the logarithm inside the $\bar{K}$
function has a negative argument for $s<0$ (since the integration variable
$s'\ge 4 M_\pi^2$). This is a generic feature of all $\bar{K}$ functions,
which is present, in general, for all values of the indices, and is the
origin of the left-hand cut. The right-hand cut is generated by the Cauchy
kernel explicitly shown in Eq.~\eqref{eq:K0000}, and which is present only
in diagonal kernels. This reflects the fact that the unitarity relation in
the physical region is diagonal: $\im t^I_\ell(s)= \sigma(s)
|t^I_\ell(s)|^2$. 

\subsection{Solving the Roy equations numerically}
In their complete form, Eq.~\eqref{eq:Royfull}, the Roy equations do not
seem to be of much use: solving an infinite system of non-linear, coupled
integral equation seems, and certainly is, prohibitive. In order to solve
them in numerical form, one needs to introduce some approximations and to
limit both the energy range and the number of waves for which one tries to
find a solution. Limiting the energy range does not mean to cut off the
integrals, of course, as this would introduce an uncontrolled systematic
uncertainty. One can rather use the fact that the $\pi \pi$ scattering
process has been measured, even though indirectly, and that the high-energy
behaviour can be theoretically controlled with the help of Regge theory.
The latter offers a parametrization, but to then use this in numerical
calculations, one needs to extract the related parameters from data in some
way.

There is also the issue of the region of validity of the Roy equations.
Since their derivation relies on the partial-wave expansion, they can only
be valid in the region where this expansion is known to converge. The
problem is, essentially, that the projection implies an integration over
$t$, and that if $s$ is beyond a certain value, the integral over $t$ will
go over the double-spectral region. In this region the analytic properties
of the full amplitude cannot be reconstructed starting from the partial
waves. The first derivation of the region of validity of the Roy equations
is due to Roy himself, who concluded that it was $-4 M_\pi^2<s<60 M_\pi^2$.
The upper limit has later been extended to $64 M_\pi^2$ and finally to
$s_1=68 M_\pi^2 \sim (1.15\; \mbox{GeV})^2$, for more details and an
account of the relevant literature see
Ref.~\cite{Ananthanarayan:2000ht,Caprini:2005zr}. 

Phenomenologically it is well known that the only partial waves which are
significantly different from zero below 1 GeV are the $S$ and $P$ waves.
Close to threshold the centrifugal barrier factor, $t \sim q^{2 \ell}$,
implies that only the $S$ waves (of both isospin 0 and 2) can be different
from zero and, for the others, that the higher the angular momentum the
stronger the suppression. Already for the $P$ wave (in this case isospin
can only be 1, as symmetry implies that even (odd) values of angular
momenta can only come in states of even (odd) isospin), the suppression
with respect to the $S$ wave extends well over the immediate neighborhood
of threshold and seems to be overcome only by the presence of the $\rho$
resonance, around $0.77$ GeV. For $D$ and higher waves the qualitative
behaviour is similar: the suppression is overcome only by the presence of a
resonance. For the $D0$ wave (the number after the letter refers to the
isospin), the first resonance occurs around $1.27$ GeV, slightly above the
limit of validity, whereas for yet higher waves the first resonances occur
even higher. So, it is reasonable to consider the imaginary parts of $D$
and higher waves as known input, because they are dominated by the
lowest-lying resonances, whose properties are reasonably well known. The
extrapolation down to threshold can also be parametrized in a simple way
and constrained within reasonable uncertainties.

In this setting all one needs to do is to consider the three $S$ and $P$
waves, namely S0, S2 and P1, and solve the Roy equations for these, up to,
maximally, $s_1$. Before attempting to solve the equations numerically, it
would be important to have some guidance for what concerns the mathematical
properties of the equations and in particular: under which conditions do
they admit a solution, or perhaps a multiplicity of solutions? Luckily
these properties have been investigated during the seventies, soon after
the equations were discovered, in particular by Wanders and
collaborators~\cite{Pomponiu:1975bi,Epele:1977um}, and revisited more
recently by Gasser and Wanders~\cite{Gasser:1999hz}. But historically, the
first numerical solutions of Roy equations were obtained with the help of
the computers available at the time, even before these mathematical properties
were established. Perhaps the most comprehensive study is the one by
Basdevant, Froggatt and Petersen, dating back to 1973~\cite{Basdevant:1973ru}.

An attentive reader will not overlook the fact that this is the same year
in which asymptotic freedom was
discovered~\cite{Gross:1973id,Politzer:1973fx}, and will probably not be
surprised by the fact that the study of these equations was essentially
abandoned after that. It took about 25 years before it was taken up again
by B.~Ananthanarayan, J.~ Gasser, H.~Leutwyler and myself. The motivation
at that time was the possibility to combine the dispersive analysis based
on Roy equations with the chiral expansion.  At that time the $\pi \pi$
scattering amplitude had been calculated at NNLO in the chiral
expansion~\cite{Bijnens:1995yn,Bijnens:1997vq}. The extensive study was
published in Physics Reports in the year
2000~\cite{Ananthanarayan:2000ht}. One year later the results of the
matching between the chiral and the dispersive representations were
published~\cite{Colangelo:2001df}, leading to a very sharp determination of
the $\pi \pi$ scattering at low energy, in particular of the two $S$-wave
scattering lengths. The prediction has been confirmed experimentally some
years later with about the same level of precision by the NA48/2
collaboration~\cite{NA482:2010dug}. 

In the following I will discuss the properties of the Roy equation
solutions obtained in Ref.~\cite{Ananthanarayan:2000ht}. The starting
point is the splitting mentioned above between the contribution of the $S$
and $P$ waves to the dispersive integrals from the rest:
\be
t^I_\ell(s)_\mathrm{SP} = k_\ell^I(s)+
\sum_{I'=0}^2\sum_{\ell'=0}^1  \int_{4M_\pi^2}^{s_2}
ds'\,K_{\ell\ell'}^{I I'}(s,s')\,\im \, t_{\ell'}^{I'}(s') \;,
\ee
with integrals cut off at $s_2$, $\sqrt{s_2}= 2$ GeV, the maximal energy on
which we have reliable phenomenological information. The remainder of the
partial wave amplitude, 
\begin{align}
\label{eq:dt} 
d_\ell^I(s)=\sum_{I'=0}^2 &\left[\; \sum_{\ell'=2}^\infty  
\int_{4M_\pi^2}^{s_2} ds'\,K_{\ell\ell'}^{I I'}(s,s')\,\im \,
t_{\ell'}^{I'}(s') \right.
 \nonumber \\ 
& \left. +\sum_{\ell'=0}^\infty  \int_{s_2}^{\infty}
ds'\,K_{\ell\ell'}^{I I'}(s,s')\,\im \, t_{\ell'}^{I'}(s') \right] \; ,
\end{align}
is called the {\em driving term}. It accounts for those contributions 
to the r.h.s.~of the Roy equations that arise from the imaginary parts of the 
waves with $\ell=2,3,\ldots$ and in addition also contains
those generated by the imaginary parts of the $S$- and $P$-waves above 2
GeV, which will be described in terms of a Regge parametrization.
By construction, we have
\be 
t^I_\ell(s)= t^I_\ell(s)_\mathrm{SP} +d^I_\ell(s) \; ,
\ee
so that this treatment of the Roy equations does not rely on any
truncation.

\begin{figure}
\includegraphics[width=6cm]{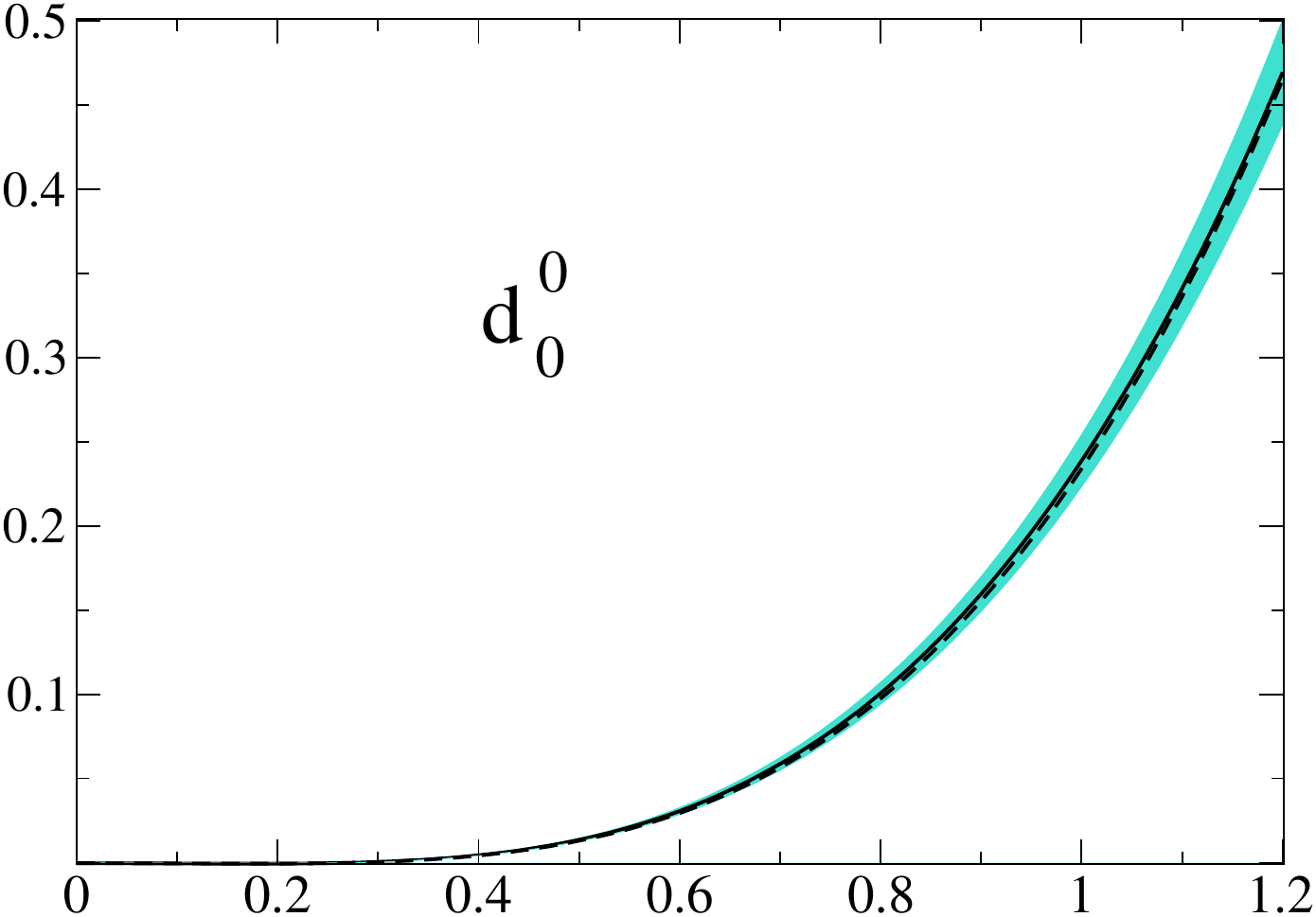}
\includegraphics[width=6cm]{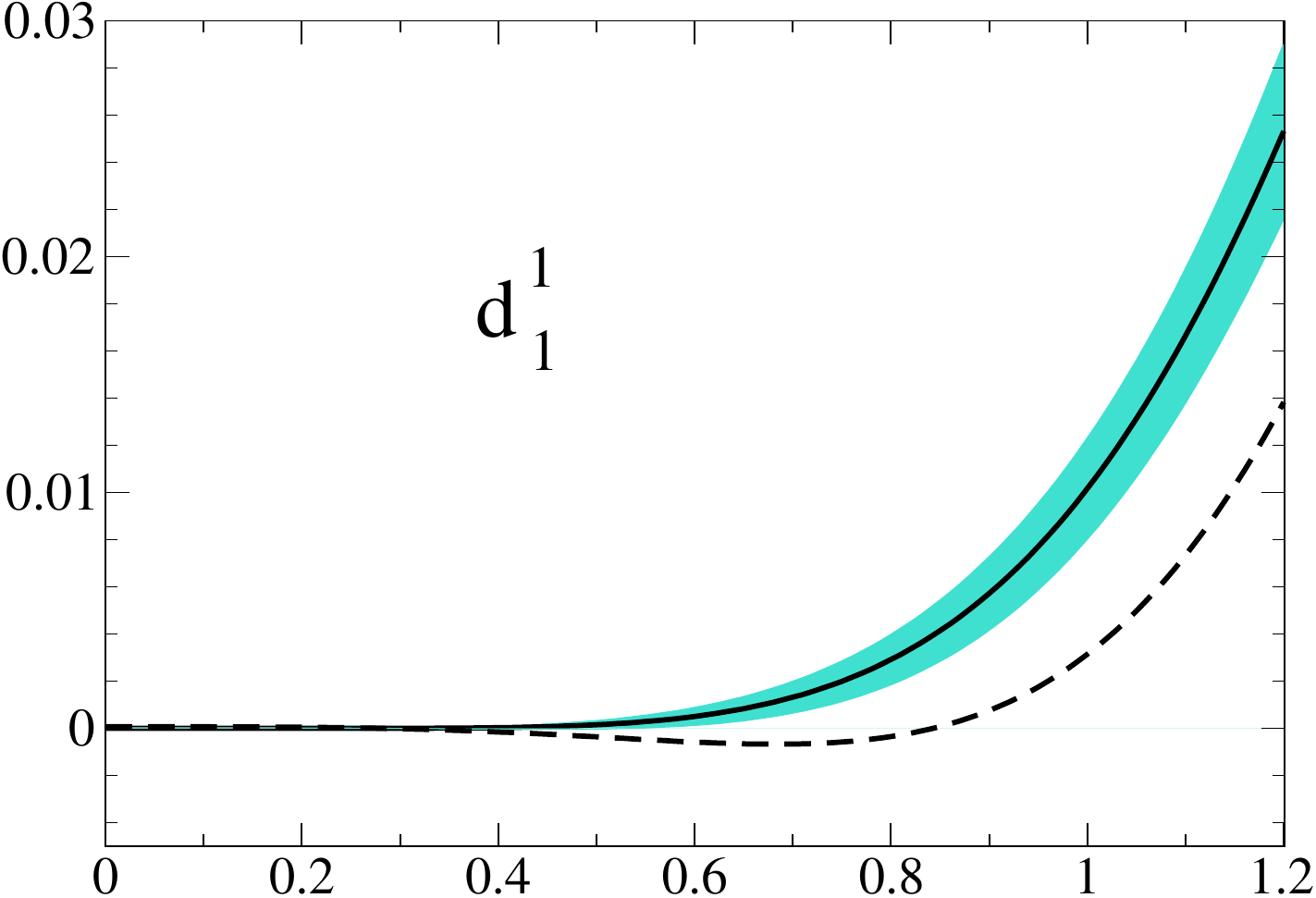}
\includegraphics[width=6cm]{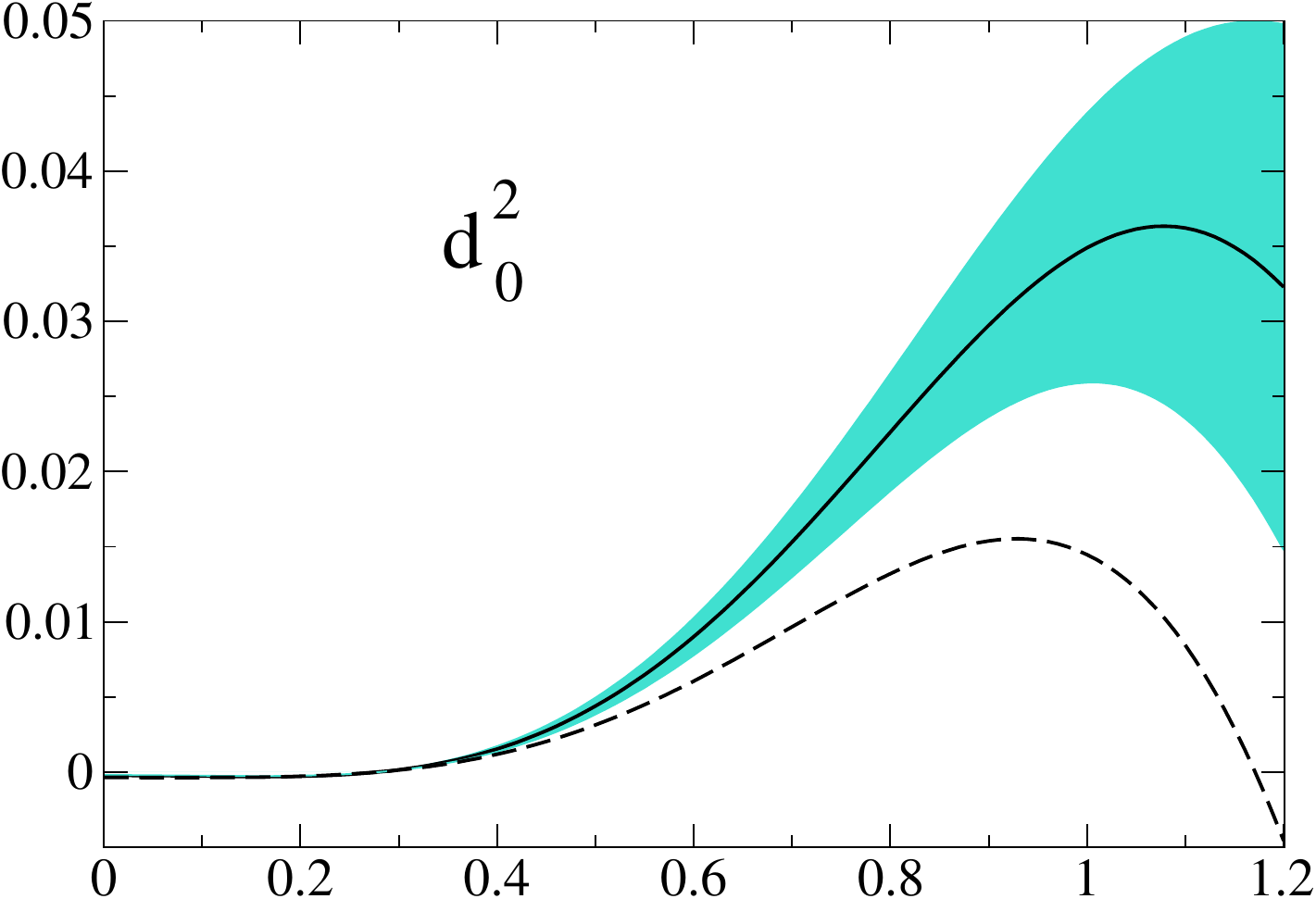}
\caption{\label{fig:driving} Driving terms versus energy in GeV. 
The full lines 
show the result of the calculation described in appendix B of
Ref.~\cite{Ananthanarayan:2000ht}. The shaded regions indicate the
uncertainties associated with the input of that calculation.  The dashed
curves represent the contributions from the $D$- and $F$-waves below 2
GeV. Notice the different scales on the $y$ axis. Figures from
Ref.~\cite{Ananthanarayan:2000ht}.} 
\end{figure}

The evaluation of the driving terms requires input about $D$ waves and
higher waves, as well as an estimate of the asymptotic amplitude, the one
which can be described by a Regge parametrization, above $s_2$. Details of
this analysis can be found in Ref.~\cite{Ananthanarayan:2000ht}. Here we
only show the plots of the driving terms in Fig.~\ref{fig:driving}

\begin{figure}[tb]
\begin{center}
\includegraphics[width=7.2cm]{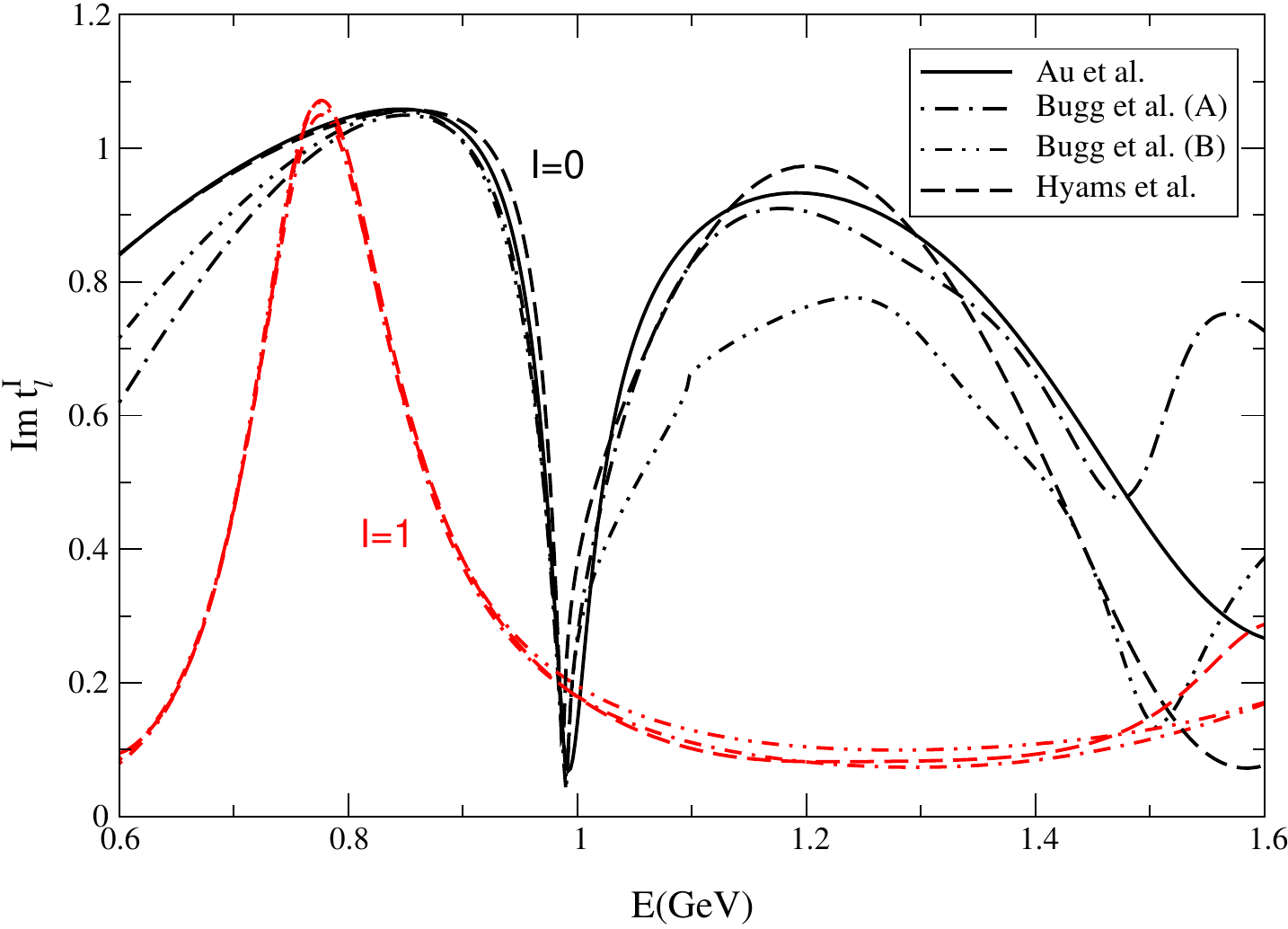}
\includegraphics[width=5.7cm]{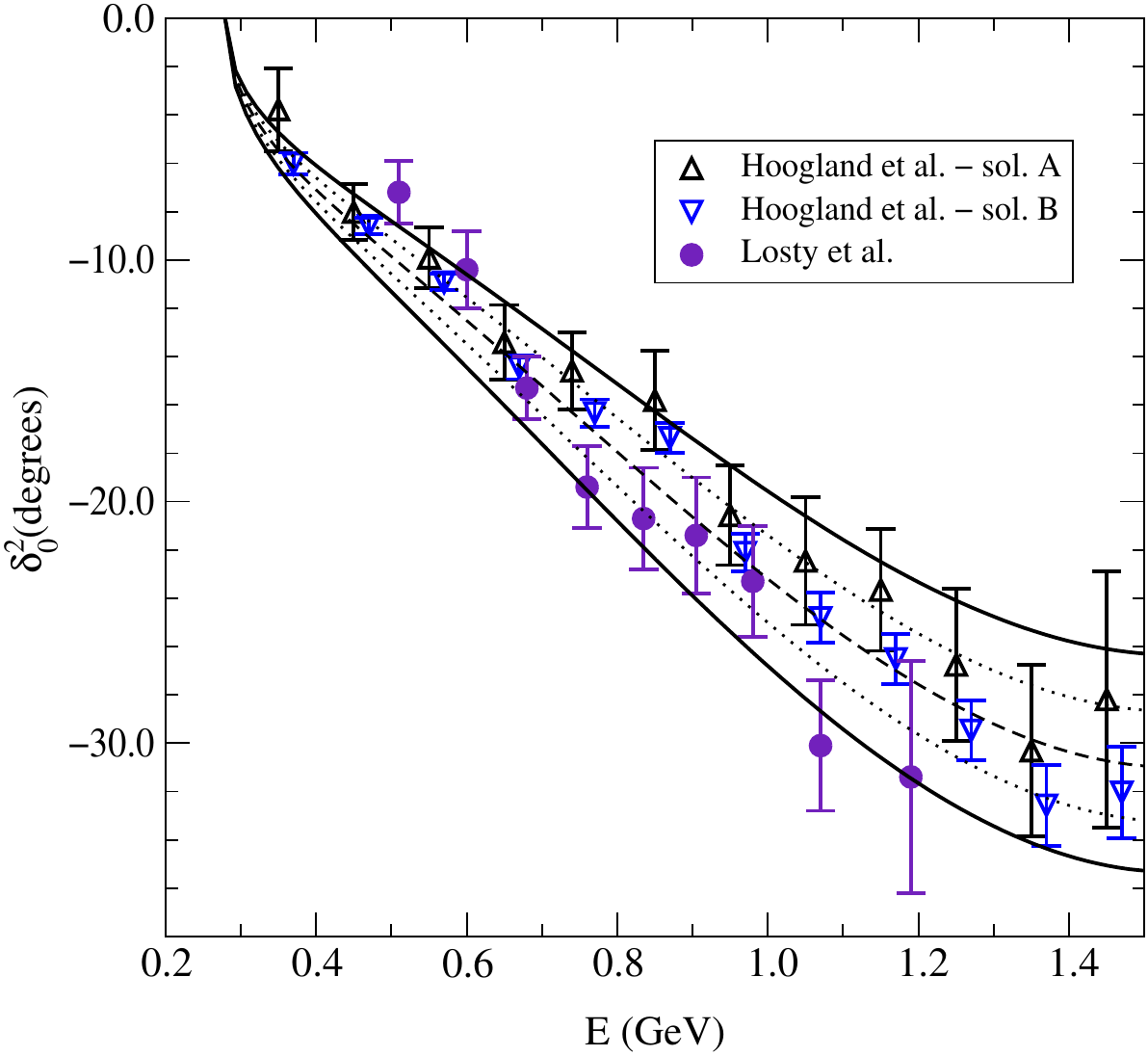}
\end{center}
\caption{\label{fig:Im012} Left: Comparison of the different input we used for the
  imaginary parts of the $I=0$ and $I=1$ lowest partial waves above the
  matching point at 0.8 GeV. Right: Different data sets for the $S2$ wave
  curves used as input in the Roy equation analysis. Figures from
  Ref.~\cite{Ananthanarayan:2000ht}.}  
\end{figure}
\begin{figure}[tb]
\begin{center}
\leavevmode
\includegraphics[width=10cm]{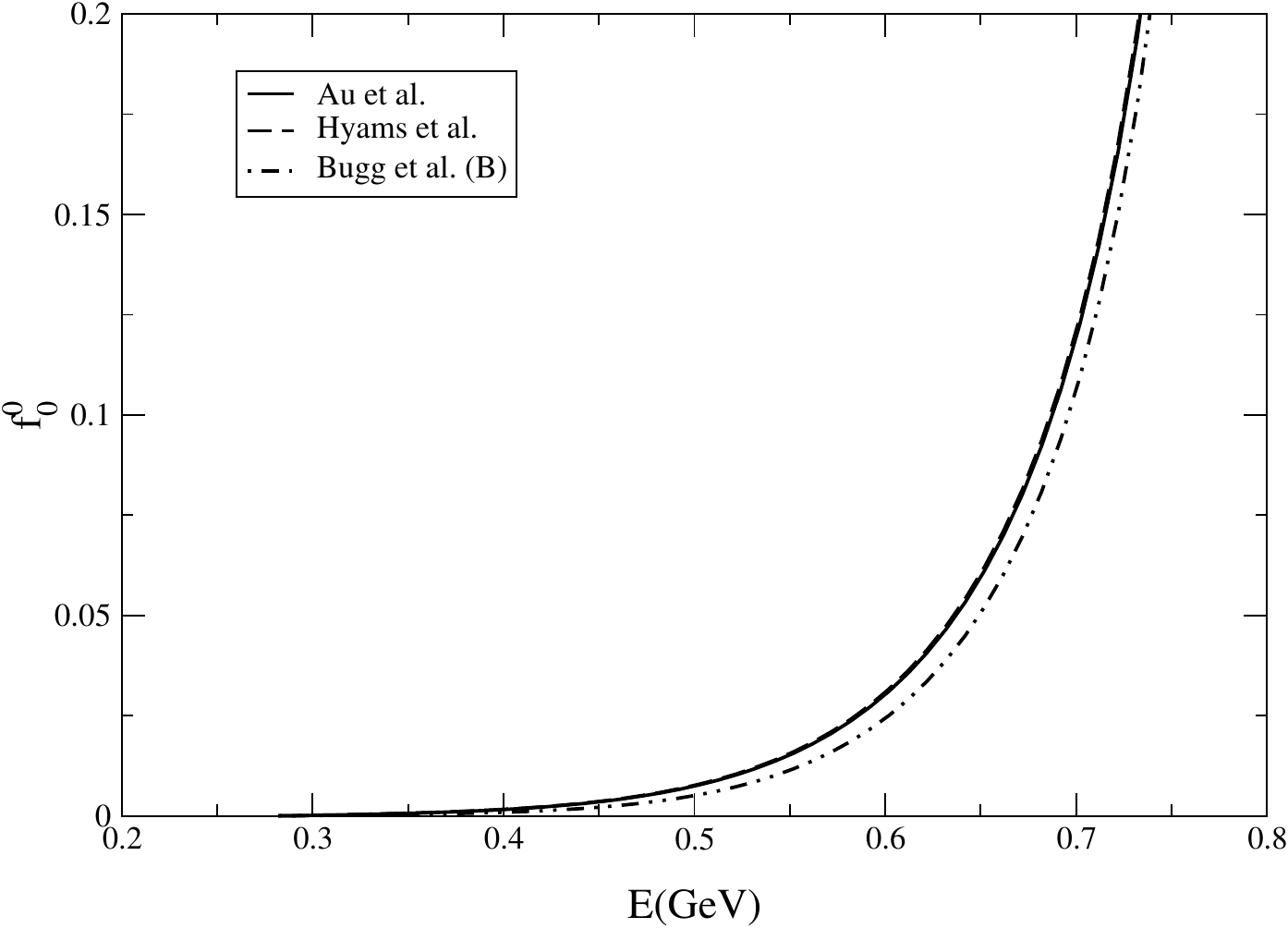}
\end{center}
\caption{\label{fig:RoyAHSZ} Comparison of the results obtained for the
  dispersion integral $f^0_0$ with the various imaginary parts
  shown in Fig.~\protect{\ref{fig:Im012}}, left panel. Figures from
  Ref.~\cite{Ananthanarayan:2000ht}.}
\end{figure}

Before solving the equations a further splitting is necessary since, as we
discussed above, the equations are not valid up to $s_2$, but maximally up
to $s_1$. Moreover, for reasons related to the multiplicity of the
solutions (see Ref.~\cite{Gasser:1999hz} and references therein), in
Ref.~\cite{Ananthanarayan:2000ht} the choice was adopted to solve the
equations only up to $\sqrt{s_0}=0.8$ GeV. This guarantees that the
solution is unique. 
According to this, the Roy equations for the $S$ and $P$ waves may be
rewritten as
\begin{align}\label{eq:rieq1}
\re\, t_\ell^I(s) &=  k_\ell^I(s)
\hspace{-1mm}+ \Pint_{4M_\pi^2}^{s_0} ds'
K_{\ell\, 0}^{I\, 0} (s,s')\, \im\, t_0^0(s')\hspace{-1mm} +
\Pint_{4M_\pi^2}^{s_0}
ds' K_{\ell\, 1}^{I\, 1} (s,s')\, \im\, t_1^1(s') \nonumber \\ 
& +
\Pint_{4M_\pi^2}^{s_0} ds' K_{\ell \, 0}^{I\, 2} (s,s')\, \im\,
t_0^2(s') + f_\ell^I(s) + d_\ell^{I}(s) \; \; ,
\end{align}
where $I$ and $\ell$ take only the values ($I, \ell$) =(0,0), (1,1)
and (2,0). The bar across the integral sign stands for principal value
integral. The functions $f_\ell^I(s)$ contain the part of the dispersive integrals
over the three lowest partial waves that comes from the region between $s_0$
and $s_2$, which is fixed in terms of experimental data as input.
They are defined as
\be
\label{eq:rieq2}
f_\ell^I(s) = \sum_{I'=0}^2\sum_{\ell'=0}^1\Pint_{s_0}^{s_2} ds' 
K_{\ell \ell'}^{I I'} (s,s')\, \im\, t_{\ell'}^{I'}(s') \fs
\ee
The input used in Ref.~\cite{Ananthanarayan:2000ht} is shown
in Fig.~\ref{fig:Im012}, which gives an idea of the related uncertainties. Note,
however, that the Roy equations are double subtracted, which means that at
low energy the influence of high-energy contributions to the integrals will
be suppressed. The relatively large variations among the possible inputs
around 1.6 GeV, is barely visible below 0.8 GeV, as Fig.~\ref{fig:RoyAHSZ} neatly
illustrates, in the case of the $S0$ wave.

\begin{figure} 
\leavevmode \centering
\includegraphics[width=13cm]{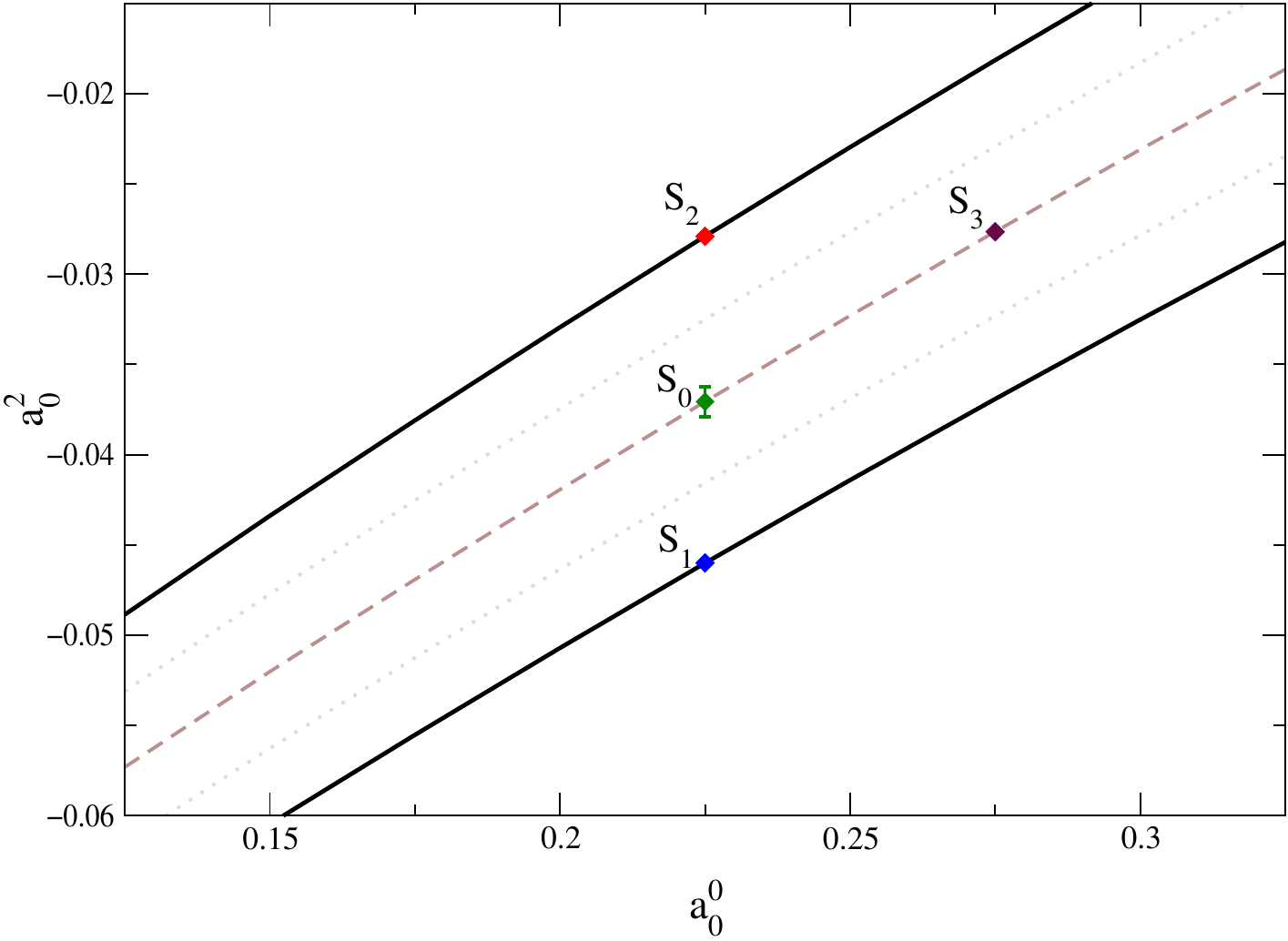}
\caption{\label{fig:UB} Universal band. The five lines
  correspond to the five different curves shown in
  Fig.~\protect{\ref{fig:Im012}} (the top line, for instance, 
results if  the input for  $\im t_0^2$ in the region above 0.8 GeV
  is taken from the top curve in that figure). 
$S_0$ marks our reference point: $a_0^2=0.225$, $a_0^2=-0.0371$. The bar
attached to it indicates the uncertainty in $a_0^2$ due to the one
in the phase $\delta_0^0$ at the matching point -- the most important
remaining source of error if the input for $\im t_0^2$ is held
fixed. Figures from Ref.~\cite{Ananthanarayan:2000ht}. }
\end{figure}

Significant uncertainties are also seen in the $S2$ wave, see
Fig.~\ref{fig:Im012}, right panel, but their impact on the solutions is
different. As it was realized already in the seventies, when the first
numerical solutions of Roy equations were found, the input for the $S2$
wave imposes a correlation between the two $S$ wave scattering lengths
$a_0^0$ and $a_0^2$: for any given input above $s_0$ for the $S2$ wave and
for a given value of $a_0^0$, there is only one value of $a_0^2$ which
gives a solution without a cusp (a discontinuity in the first derivative)
at $s_0$ in the $S2$ wave. To each of the curves in the right panel of
Fig.~\ref{fig:Im012} corresponds a line in the $(a_0^0,a_0^2)$ plane. Since
the upper and lower curves in the right panel of Fig.~\ref{fig:Im012}
roughly provide upper and lower bounds of the possible phenomenological
inputs for this wave, the corresponding curves in the $(a_0^0,a_0^2)$ plane
delimit the region where it is possible to find physical solutions of the
equations (given the available phenomenological input). The region of the
$(a_0^0,a_0^2)$ plane in which physical solutions are possible is called
the ``Universal band'' and is shown in Fig.~\ref{fig:UB}.
In this figure four different points are (arbitrarily) singled out, just to
illustrate how the solutions look like and how they differ if the
scattering lengths are changed, while the rest of the phenomenological
input is held fixed. 
\begin{figure} 
\leavevmode \centering
\includegraphics[width=8cm]{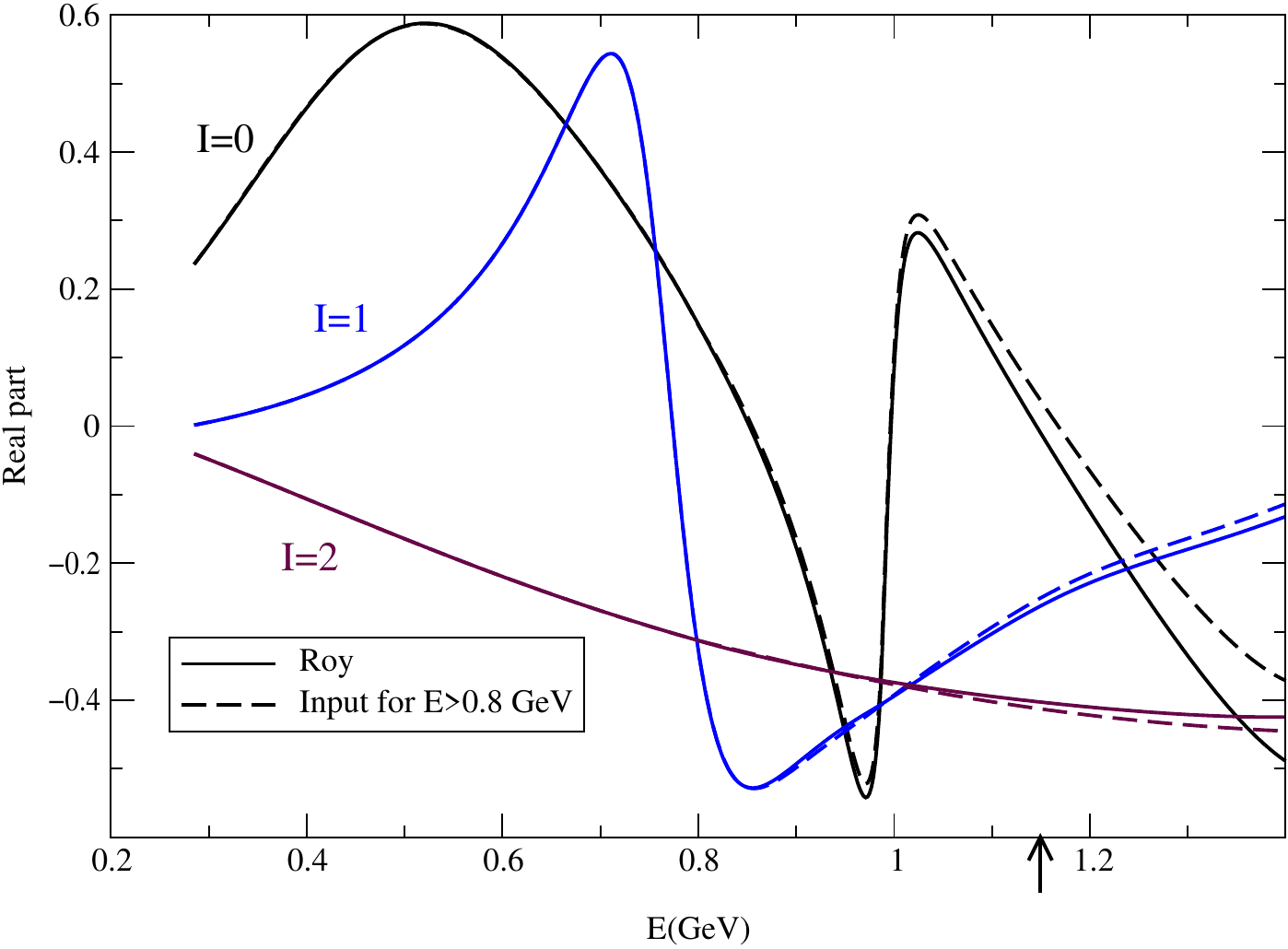}
\includegraphics[width=8cm]{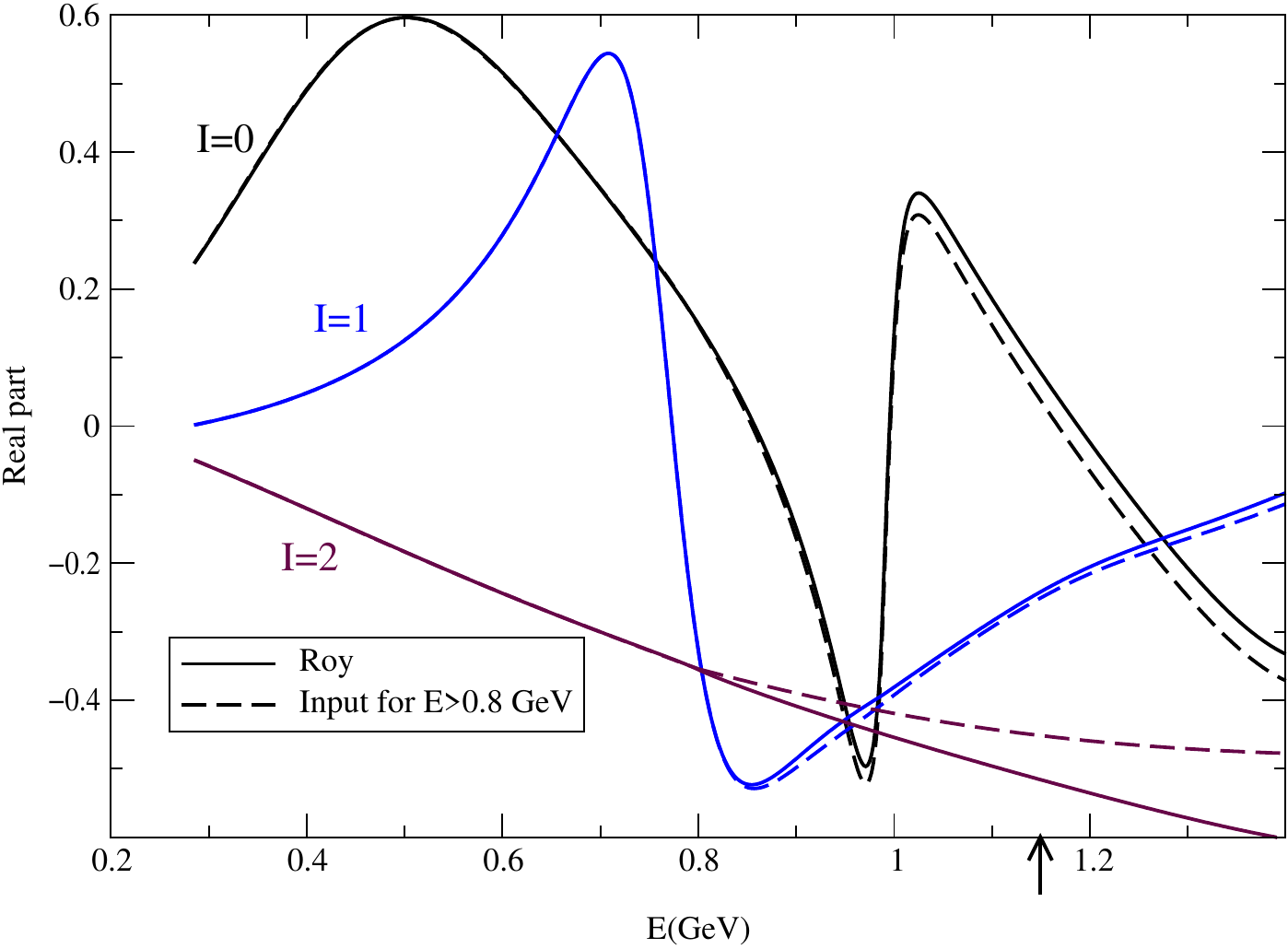}
\includegraphics[width=8cm]{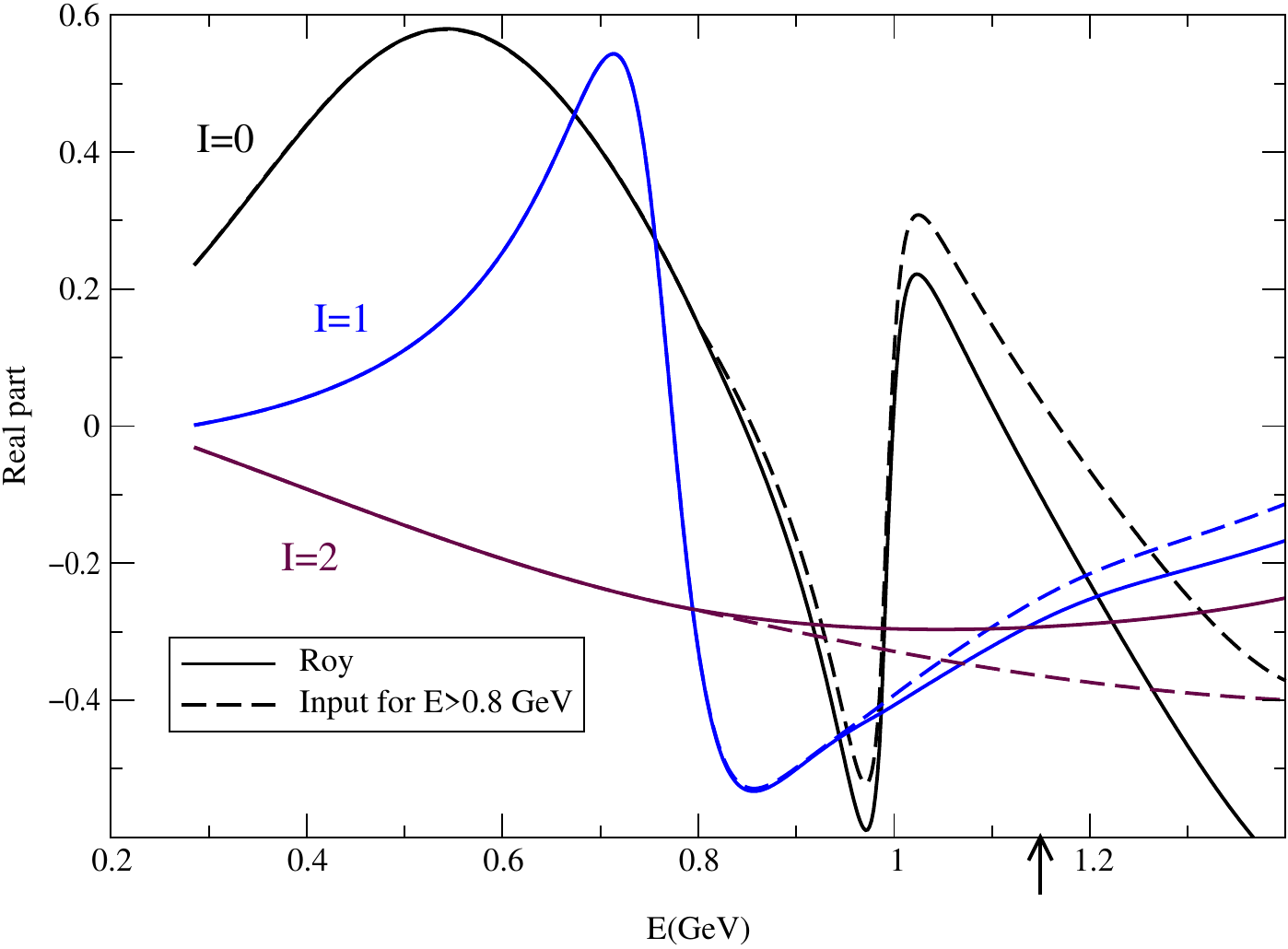}
\caption{\label{fig:RoyS012} Numerical solution of the Roy equations for
$a_0^0=0.225$, $a_0^2=-0.0371$, point S$_0$ in Fig.~\ref{fig:UB}, upper
panel. The middle and lower panels are numerical solutions for points S$_1$
and S$_2$ respectively (the value of $a_0^0$ corresponds to the center
of the range considered while the one of $a_0^2$ results if 
the input used for $\mbox{Im}\,t^2_0$ is taken from the  central, lower, or
upper curve in Fig.~\protect{\ref{fig:Im012}}, right panel). The arrow indicates the limit
of validity of the Roy equations. Figures from Ref.~\cite{Ananthanarayan:2000ht}. }
\end{figure}
The solutions corresponding to points S$_0$, S$_1$ and
S$_2$ are shown in Fig.~\ref{fig:RoyS012}. As one can see there, the
difference between the left- and the right-hand side of the Roy equations
is invisible below $s_0$, the region where the equations have actually been
solved. Above that point both curves are continued to be shown, with
differences which become more and more evident as the energy grows. It is
quite remarkable that, given the phenomenological input described above,
what happens between $s_0$ and threshold is completely determined by the
values of the two scattering lengths, and the mathematics of these integral
equations, of course.

In those plots the limit of validity of the Roy equations is explicitly
shown, which brings to mind the question whether it would be possible to
find solution all the way up to $s_1$. This has actually been done by
I.~Caprini, H.~Leutwyler and myself in Ref.~\cite{Caprini:2011ky}. There
are a number of complications linked to this extension:
\begin{enumerate}
\item
inelasticities start in principle at $16 M_\pi^2$, but effectively they are
still compatible with zero up to $s_0$. Solving the equations up to $s_1$
requires an input for the inelasticities in all channels;
\item
the solution of the Roy equations is not unique any more. This means that
even if one has reduced the energy region in which phenomenological input
is required (pushing up the lower boundary from 0.8 to 1.15 GeV), one still
needs some phenomenological input below $s_1$. As discussed in
Ref.~\cite{Caprini:2011ky} one now needs three new inputs to pin down the
physical solution within the three-dimensional manifold of possible
mathematical solutions. 
\item
the simple Schenk parametrization of the phase-shifts adopted in
Ref.~\cite{Ananthanarayan:2000ht} to look for solutions is not adequate any
more in the broader range going up to $s_1$. A more complicated
parametrization with several more parameter has been used in
Ref.~\cite{Caprini:2011ky}.
\item
extending the range up to $s_1$ implies that both the resonances in the $D$
and higher waves, as well as the Regge parametrization have a more
significant impact on solutions. To reduce uncertainties, the Roy equations
for $D$ and $F$ waves have been solved and the Regge parameters have been
constraint to satisfy a set of sum rules which follow from the requirement
of full crossing symmetry---a property not automatically satisfied by
Roy-equation solutions.
\end{enumerate}
The interested reader is referred to~\cite{Caprini:2011ky} for more
details. 

So what are these solutions good for? As we have seen, if one assumes that
all the phenomenological input needed is correct (within uncertainties),
then the Roy equations tell us that at low energy, the essential parameters
are the two scattering lengths. The closer to threshold one measures the
$\pi \pi$ scattering amplitude, the stronger the sensitivity to these
parameters. Any measurement of $\pi \pi$ scattering in the low-energy
region is therefore a measurement of the two $S$-wave scattering
lengths. And given the fact that the Roy equations just follow from
analyticity, unitarity and crossing symmetry, one can use these solutions
to fit the data, using as fitting parameters just the scattering
lengths, knowing that one is not introducing any theoretical bias. 
We will come to this use of the Roy-equation solutions in the next
section. 

But before doing this I come back to the fits of the $e^+e^- \to
\pi^+\pi^-$ data discussed in the previous section and explain their logic
on the basis of what I have discussed here. I will assume for the moment
that the two scattering lengths have been fixed---as discussed in the next
section they have been predicted to high accuracy and the prediction has
been experimentally confirmed. So what free parameters do we have to fit
the $e^+e^-$ data? As I have discussed above, the Roy equations need a
phenomenological input above $s_0$, or, in case one looks for a solution in
the whole range of validity, at $s_1$ and above. Let's consider the latter
scenario. The imaginary part of the $P1$ wave used as input in
Ref.~\cite{Ananthanarayan:2000ht} is shown in Fig.~\ref{fig:Im012}. It's a
rather boring function between $s_0$ and $s_1$ and also above. In the
extended analysis discussed in~\cite{Caprini:2011ky} the difference in the
phenomenological input are not particularly significant: it is about the
same as what is shown in Fig.~\ref{fig:Im012}. The most relevant
phenomenological input, however, is the value of the phase at the matching
point: $\delta_1^1(s_0)$ in case of the small-range analysis, or
$\delta_1^1(s_1)$ in case of the extended analysis. In the latter case,
however, the multiplicity of solutions forces us to take another input even
below $s_1$. The choice adopted in~\cite{Caprini:2011ky} is to take
$\delta_1^1(s_0)$ as second input, as this makes a comparison with earlier
work much simpler. Having fixed the scattering lengths these two now are
the essential parameters for the $P1$ wave. Fixing these two, the behaviour
of the phase-shift between threshold and $s_0$ and between $s_0$ and $s_1$
is determined by the mathematics of Roy equations. Conversely, one can
calculate the phase-shift $\delta_1^1(s)$ for different values of the input
parameters $\delta_1^1(s_{0,1})$, evaluate the corresponding Omn\`es
function, and then fit the data with $\delta_1^1(s_0)$ and
$\delta_1^1(s_1)$ as free parameters. The fit outcome is shown in
Table~\ref{tab:FinalFitsSingleExperiments} and shows that these two
parameters can be determined with an astounding accuracy: better than 1\%
for $\delta_1^1(s_0)$ and better than 2\% for $\delta_1^1(s_1)$. This is
mainly the consequence of the precision measurements of the $e^+e^-$ cross
section, but also shows that, despite the compexity of the whole machinery
of the Roy equations, and the need for phenomenological input in different
waves and energy regions, this is a precision tool, in particular after
the matching with the chiral representation.

\subsection{Matching the dispersive and the chiral representations}
The simplest way in which one can think about the complementarity between
the dispersive and the chiral approach to $\pi \pi$ scattering, and how to
exploit it in terms of a matching procedure is the following. In the Roy
equations the two $S$-wave scattering lengths are the essential parameters
at low energy; in the chiral approach these can be calculated. If one
takes the values of the scattering lengths from \chpt and uses them in the
Roy equations, one has fixed the two essential and free parameters of the
dispersive approach and has a predictive framework. From the point of view
of the chiral approach this represents a significant advantage, because the
dispersive representation is valid well beyond the region of validity of
\chpt. 

\begin{figure}[t]
\centering
\includegraphics[width=12.8cm]{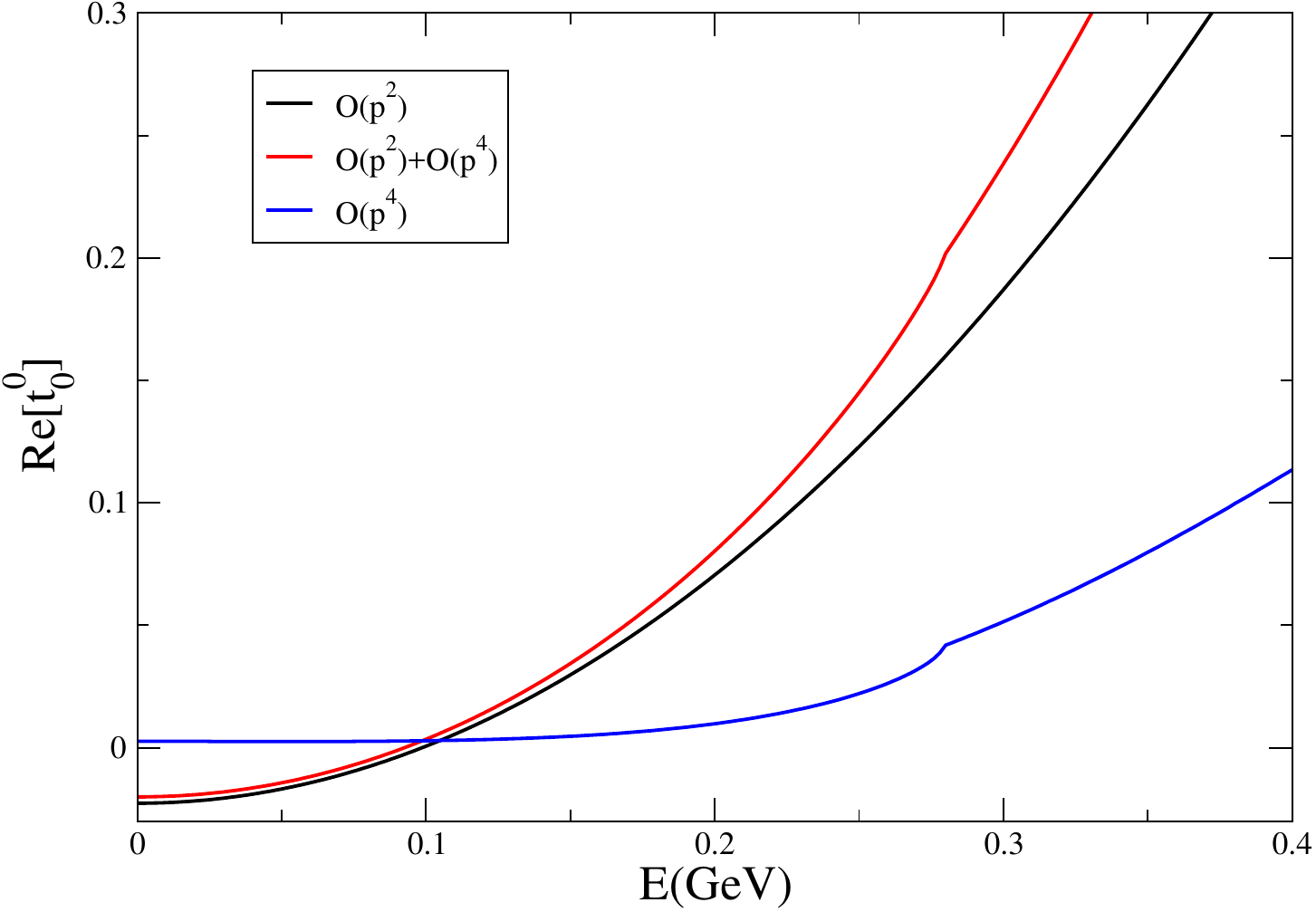}
\caption{Real part of the $S0$ partial wave in \chpt to LO, $\cO(p^2)$, and
  NLO, $\cO(p^4)$.
\label{fig:Ret00}}
\end{figure}

While this way of arguing is logically correct, there are better ways to
make it work in practice, while adhering to the same logic. The idea and
its implementation has been discussed in~\cite{Colangelo:2001df} and can be
best illustrated with the help of a plot, shown in Fig.~\ref{fig:Ret00}.
At $\cO(p^2)$, or LO, in \chpt, the $S0$ partial wave is a simple
polynomial:
\be
t_0^0(s)=\frac{2s-M_\pi^2}{32 \pi F_\pi^2} \; ,
\ee
which, evaluated at threshold gives $a_0^0=7 M_\pi^2/(32 \pi
F_\pi^2)=0.16$. At $\cO(p^4)$, or NLO, the imaginary part starts and the
real part now contains unitarity effects: contributions which display a cut
as prescribed by unitarity and analyticity. The expression for $t_0^0(s)$
is known analytically~\cite{Gasser:1983yg}, but there is no point in
showing it explicitly here: the corresponding curve is shown in
Fig.~\ref{fig:Ret00}, and what we learn from the plot is that this is a
tiny correction up to about $0.2$ GeV, but quickly gets steam as we move
towards threshold, where it becomes a 20\% correction to LO~\cite{Gasser:1983yg}:
$a_0^{0,\mathrm{NLO}}=0.20$. This shows that a matching between the chiral
and dispersive representation at threshold is not a good idea, and is by no
means a mandatory choice: the Roy equations can be subtracted at a
different point, for example at $s=0$, where the chiral expansion converges
much better, as shown in Fig.~\ref{fig:Ret00}. After doing the matching in
this way, at a subthreshold point (its exact value is not essential, as
long  as one is far enough from the threshold), one can use the dispersive
representation to reach the threshold and calculate the scattering
lengths. The improvement in convergence is stunning, as shown by the
following numbers obtained with a subthreshold matching with \chpt at LO,
NLO and NNLO:
\begin{align}
a_0^0&= 0.197 \to 0.2195 \to 0.220 \nonumber \\
-10 \cdot a_0^2&= 0.402 \to  0.446\;\, \to 0.444 \; ,
\end{align}
as compared to the same numbers evaluated directly in \chpt at LO, NLO and
NNLO: 
\begin{align}
a_0^0&= 0.159 \to 0.200 \to 0.216 \nonumber \\
-10 \cdot a_0^2&= 0.454 \to  0.445\;\, \to 0.445 \; .
\end{align}
The expectation that yet higher orders in the chiral expansion will be
irrelevant is based on this improved pattern of convergence obtained
thorugh the matching.

A detailed error analysis presented in~\cite{Colangelo:2001df} has led to
the final results
\be
a_0^0=0.220(5) \qquad a_0^2=-0.0444(10) \; ,
\ee
showing a remarkable precision at the few-percent level, which is quite
exceptional in hadronic physics. Note that, as argued above, the
uncertainties also cover possible higher-order effects.  

\begin{figure}
\centering
\includegraphics[width=6.2cm]{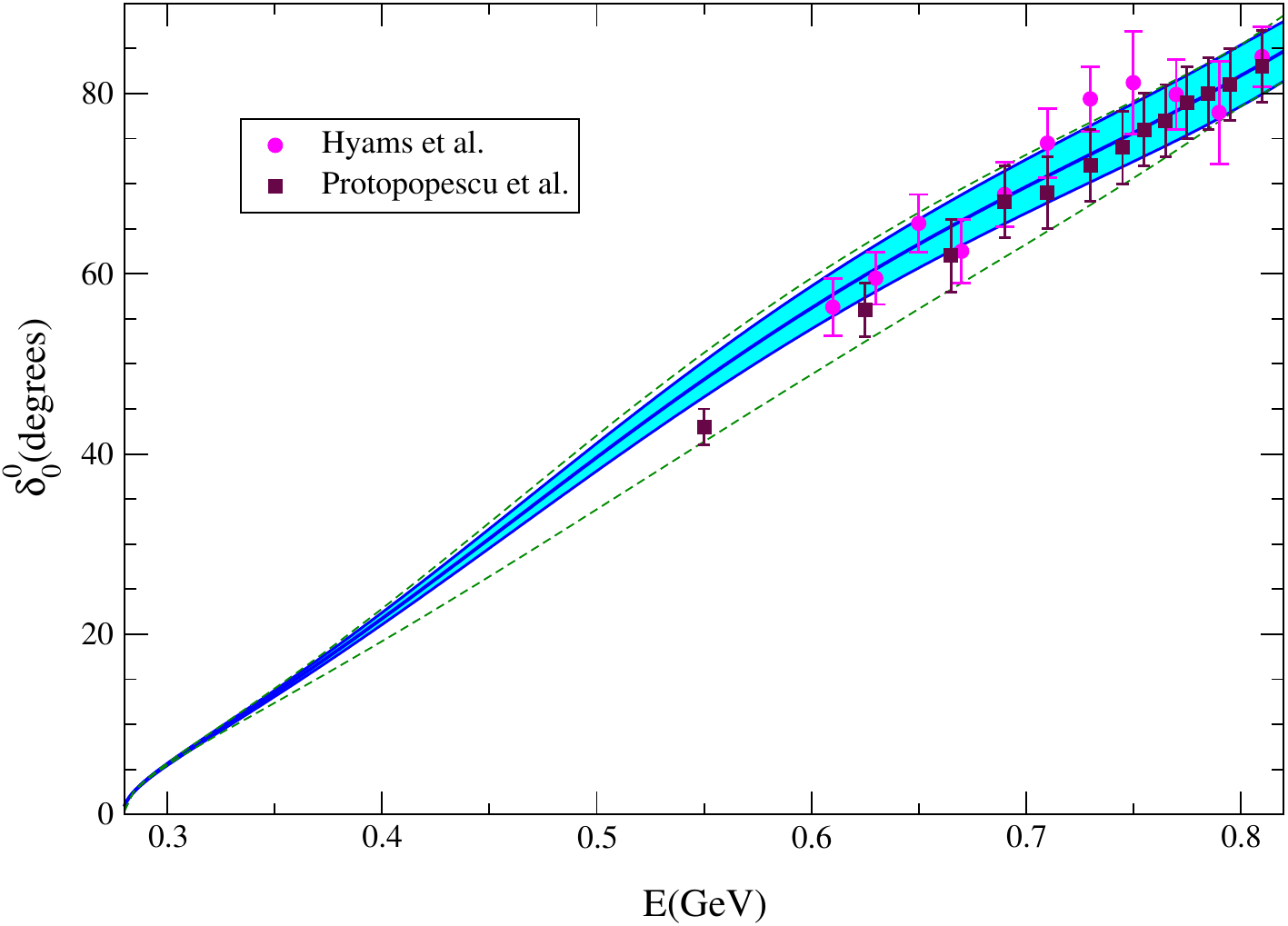}
\includegraphics[width=6.2cm]{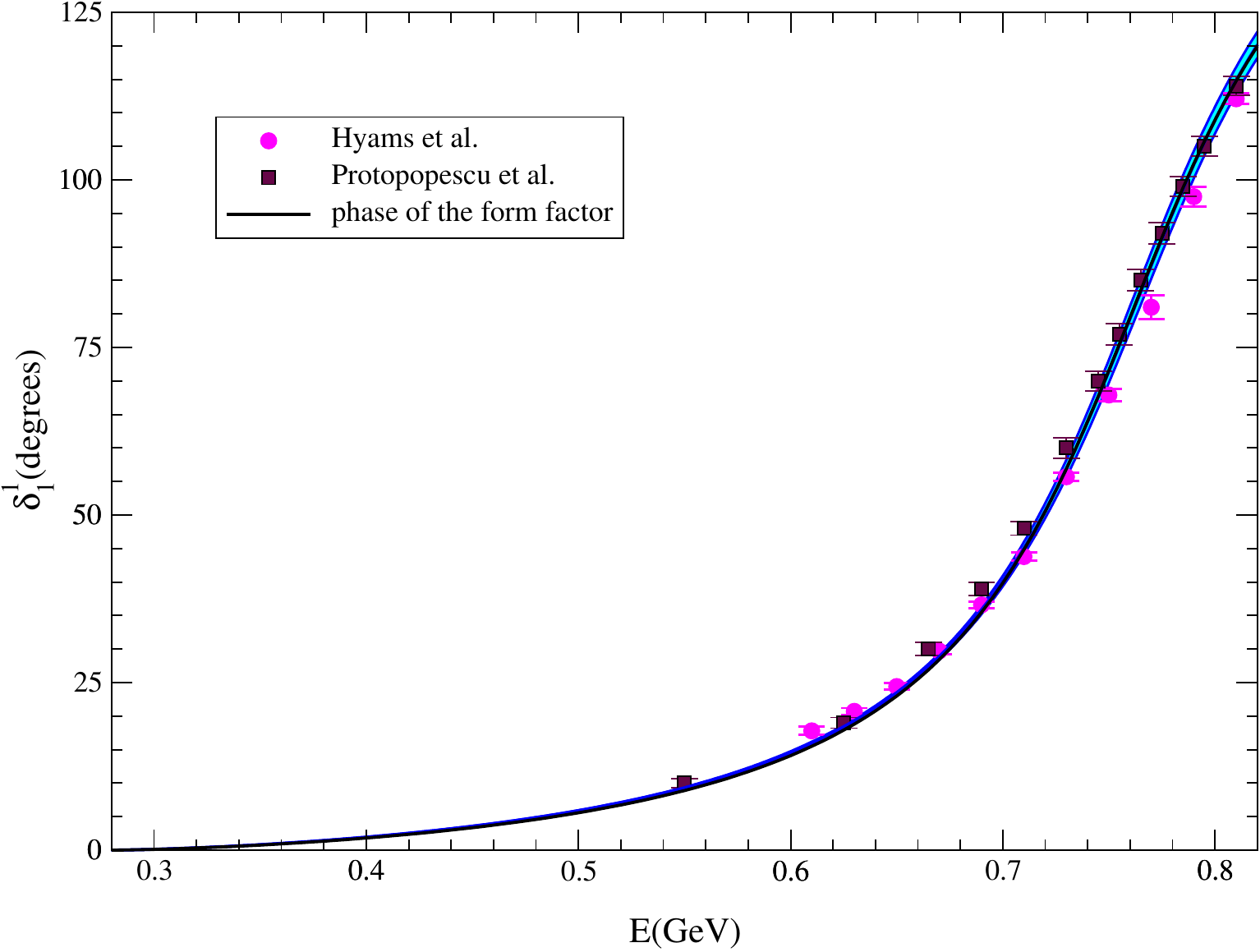}
\caption{Phase-shift for the $S0$ (left) and $P1$ (right) waves obtained after the
  matching with \chpt below threshold. In the plot for $\delta_0^0$ the
  dashed lines show the uncertainty band obtained with the Roy equations
  only, before input from \chpt. Figures from
  Ref.~\cite{Colangelo:2001df}. \label{fig:S0P1}}
\end{figure}

The impact of this matching procedure is evident also at higher energies,
as illustrated in Fig.~\ref{fig:S0P1}. In particular the plot of the $S0$
wave shows the impact of the matching with \chpt on the uncertainty: at the
matching point there is obviously no difference, but as we move down in
energy towards threshold, the reduction of the uncertainty due to chiral
constraints becomes evident.

\begin{figure}[thb]
\includegraphics[width=6.6cm]{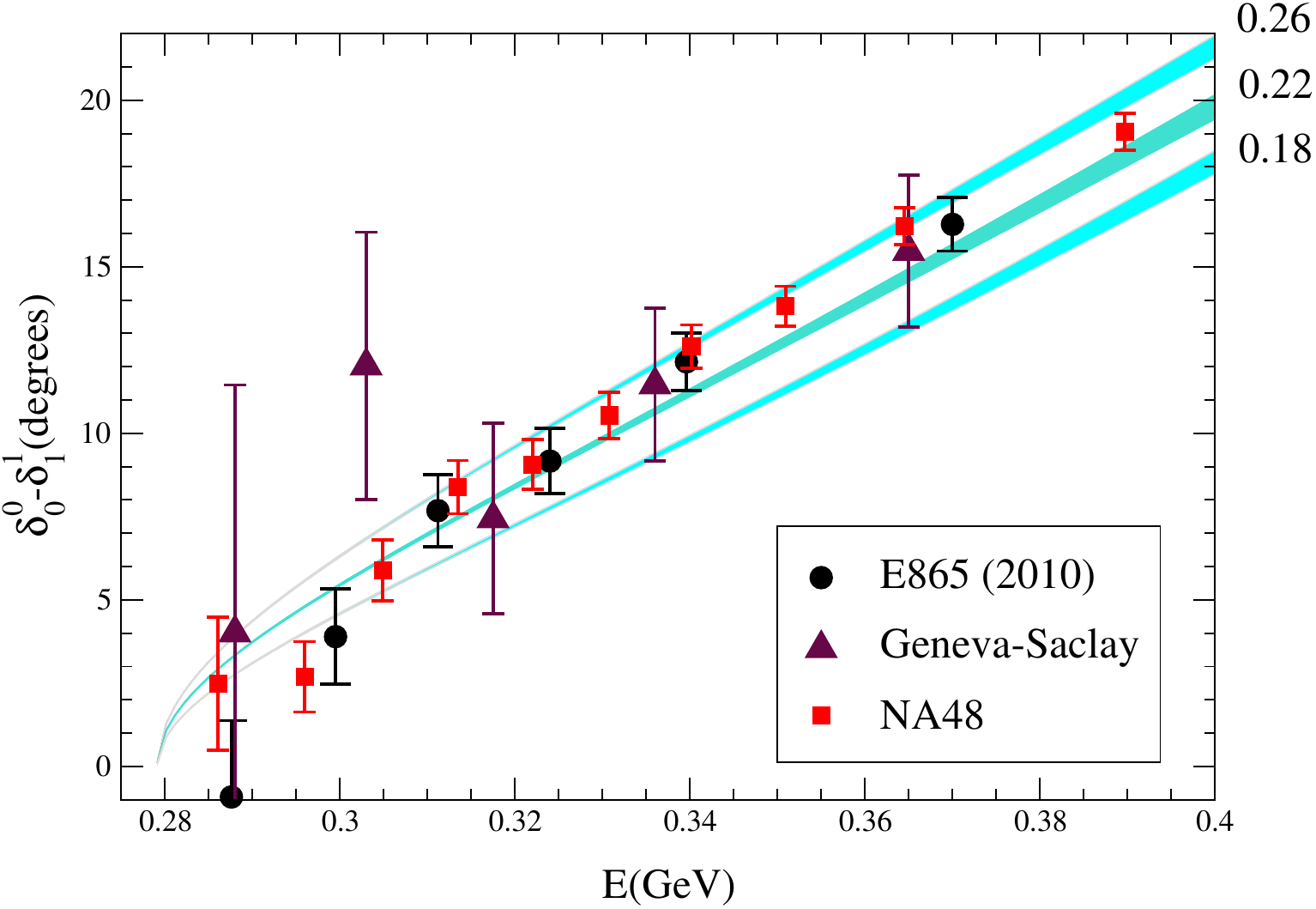}
\includegraphics[width=6.3cm]{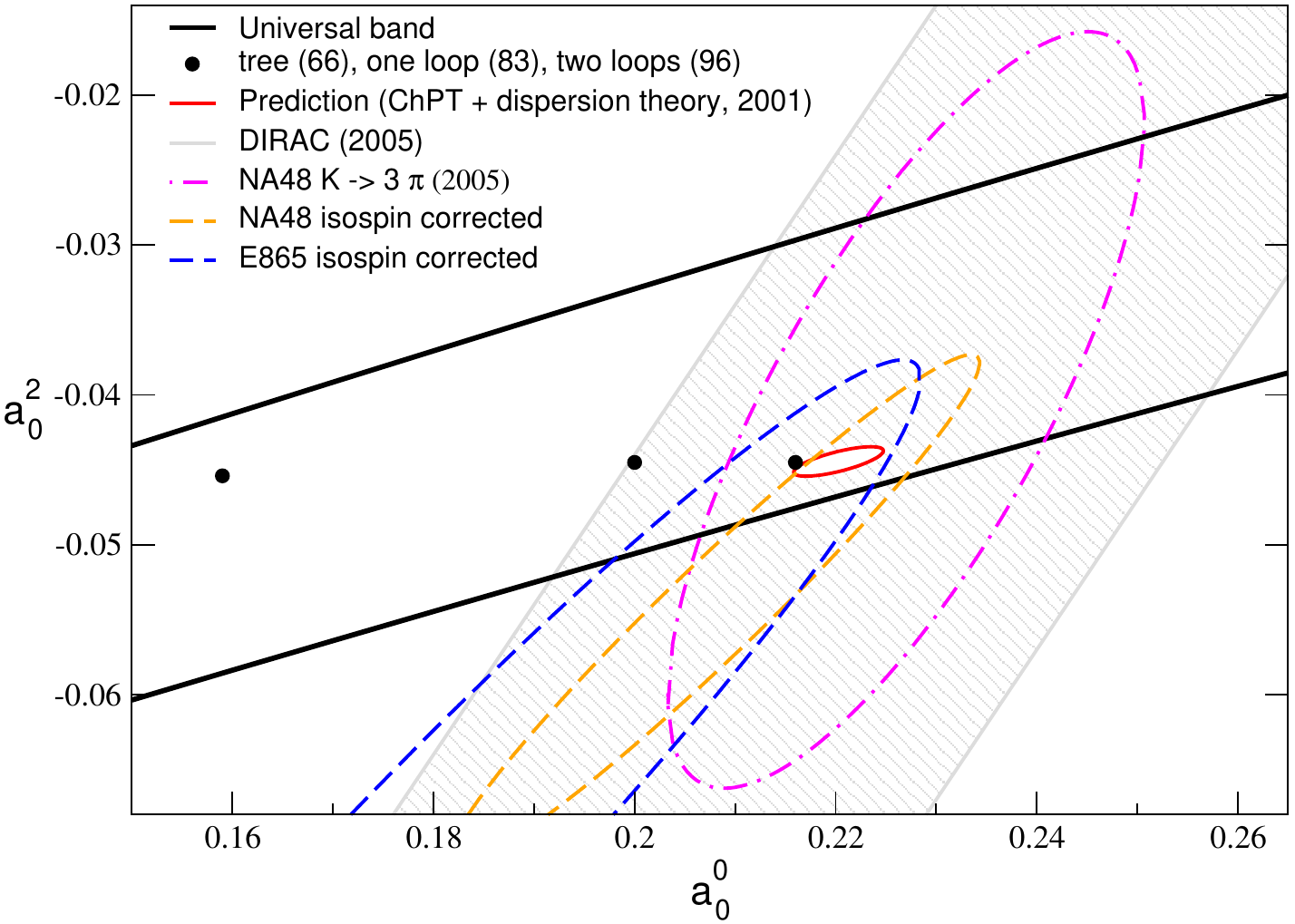}
\caption{Left: data points of the measurement of $\delta_0^0-\delta_1^1$
  performed by the Geneva-Saclay, E865 and NA48/2 collaborations, compared
  to the curves corresponding to Roy-equation solutions for different
  values of $a_0^0$. Right: comparison of the theoretical prediction of the
  two scattering lengths with the four different experimental
  determinations discussed in the text. Figures from
  Ref.~\cite{Colangelo:2008sm}. 
\label{fig:a0a2-exp}}
\end{figure}This prediction has later been confirmed experimentally through four
different, independent measurements. Two of them concerned the measurement
of the phase difference $\delta_0^0(s)-\delta_1^1(s)$ from the rescattering
of pions in $K_{e4}$ decays, so from threshold up to somewhat below the
kaon mass. This measurement has been performed both by the E865
collaboration at Brookhaven~\cite{BNL-E865:2001wfj,Pislak:2003sv} and by
NA48/2 at CERN~\cite{NA482:2010dug}. Their data and a comparison to the
Roy-equation solutions for different values of $a_0^0$ are shown in
Fig.~\ref{fig:a0a2-exp}. It turns out that at such level of precision it is
crucial to take into account isospin breaking effects in $K_{e4}$
decays~\cite{Colangelo:2008sm}. The one-$\sigma$ ellipses shown in
Fig.~\ref{fig:a0a2-exp} (right panel) take into account these effects. 

Another measurement concerns the cusp at $2 M_{\pi^+}$ in the $\pi^0 \pi^0$
invariant mass spectrum in $K^\pm \to \pi^\pm \pi^0 \pi^0$ decays, which has
been measured very precisely by the NA48/2 collaboration. Detailed
theoretical studies of these cusp effects have shown that it is possible to
accurately extract the $a_0^0-a_0^2$ difference, as the strength of the
cusp is proportional to its modulus squared (modulo small corrections which
have been estimated, and which also depend on the scattering lengths), see
Refs.~\cite{Cabibbo:2004gq,Cabibbo:2005ez,Colangelo:2006va,Bissegger:2008ff}.
The outcome of the measurement, analyzed with the theoretical formulae just
mentioned, can also be represented as a one-$\sigma$ ellipse in the
$(a_0^0,a_0^2)$ plane, which is also shown in Fig.~\ref{fig:a0a2-exp}
(right panel).

Finally, the third mesurement is that of the lifetime of pionium, which
has been performed by the DIRAC collaboration at
CERN~\cite{DIRAC:2005hsg}. The decay width is, like the cusp, also
proportional to the modulus squared of the $a_0^0-a_0^2$ difference, and
its measurement leads to a direct determination thereof. Also in this case
the theoretical apparatus needed to extract the scattering-length
difference after removing all isospin-breaking effects is quite complex and
has been developed in a beautiful series of papers by J.~Gasser,
V.~Lyubovitskij and A.~Rusetsky~\cite{Gasser:1999vf,Gasser:2001un} (see
also~\cite{Gasser:2007zt} for a general review of the theoretical framework
developed to study hadronic atoms in QCD+QED). The result of this
remarkable measurement and of the corresponding determination of
$a_0^0-a_0^2$ is also shown in Fig.~\ref{fig:a0a2-exp} (right panel).

\subsection{The $\sigma$ resonance}
Having determined the low-energy $\pi \pi$ scattering amplitude to a very
high precision is not only important for its own sake, but is a knowledge
which can be applied in several different contexts, as the example of the
HVP contribution to $(g-2)_\mu$ shows. A particularly relevant one is that
of spectroscopy of QCD, which might seem surprising, as this is considered
the realm of lattice QCD. Indeed this approach has no competition if one
wants to calculate, e.g., the proton mass from first principles. But as
soon as we consider unstable particles in QCD, i.e.  resonances, then even
lattice QCD is faced with serious challenges. The problem is that
resonances are poles of the $S$-matrix on the second Riemann sheet of the
complex plane. Evaluating on the lattice the $S$-matrix for a given real
value of its argument in the elastic region is already a highly non-trivial
problem, which was solved by L\"uscher about 40 years
ago~\cite{Luscher:1986pf}. But using information on the real axis to
determine the position of poles on the second Riemann sheet, presents both
phenomenologists as well as lattice practitioners with the same
difficulties: a pure lattice approach does not offer any advantage in this
regard. 

A dispersive representation of the scattering amplitude, and therefore of
the $S$-matrix, does, however, offer a solution to this problem. The key
observation is that for the $S$-matrix poles on the second Riemann sheet
correspond to zeros on the first, which follows from this relation between
the $S$-matrix element of angular momentum $\ell$ on the second and first
Riemann sheet:
\be
S_\ell(s)^{II}=1/S_\ell(s)^{I} \; ,
\ee
where the superscript refers the the Riemann sheet.

This was well known in the sixties, but went somewhat forgotten for some
decades, otherwise one cannot understand the debate about the existence of
the $\sigma$ resonance. Let us consider the partial wave relevant for this
resonance, the $S0$, and write the $S$-matrix element as: 
\be 
S_0^0(s)=1+ 2 i \sigma(s) t_0^0(s) \; .
\label{eq:SIvsII}
\ee
The dispersive representation for $t_0^0(s)$ which follows from the Roy
equations is valid also for complex values of its argument. The dependence
on $s$ is explicit both in the subtraction polynomial as well as in the
dispersive integrals. When evaluating the latter for complex values of $s$
one still needs as input the imaginary parts of the partial waves {\em on
  the real axis} only! Of course when doing this, one is still confined to
the first Riemann sheet, where no poles can occur. But
relation~\eqref{eq:SIvsII} tells us that to determine the position of poles
on the second Riemann sheet I just need to look for zeros on the first, and
having a reliable representation of the imaginary parts of the relevant
partial waves {\em on the real axis} is all one needs to perform this
search. 

\begin{figure}[t]
\includegraphics[width=6.4cm]{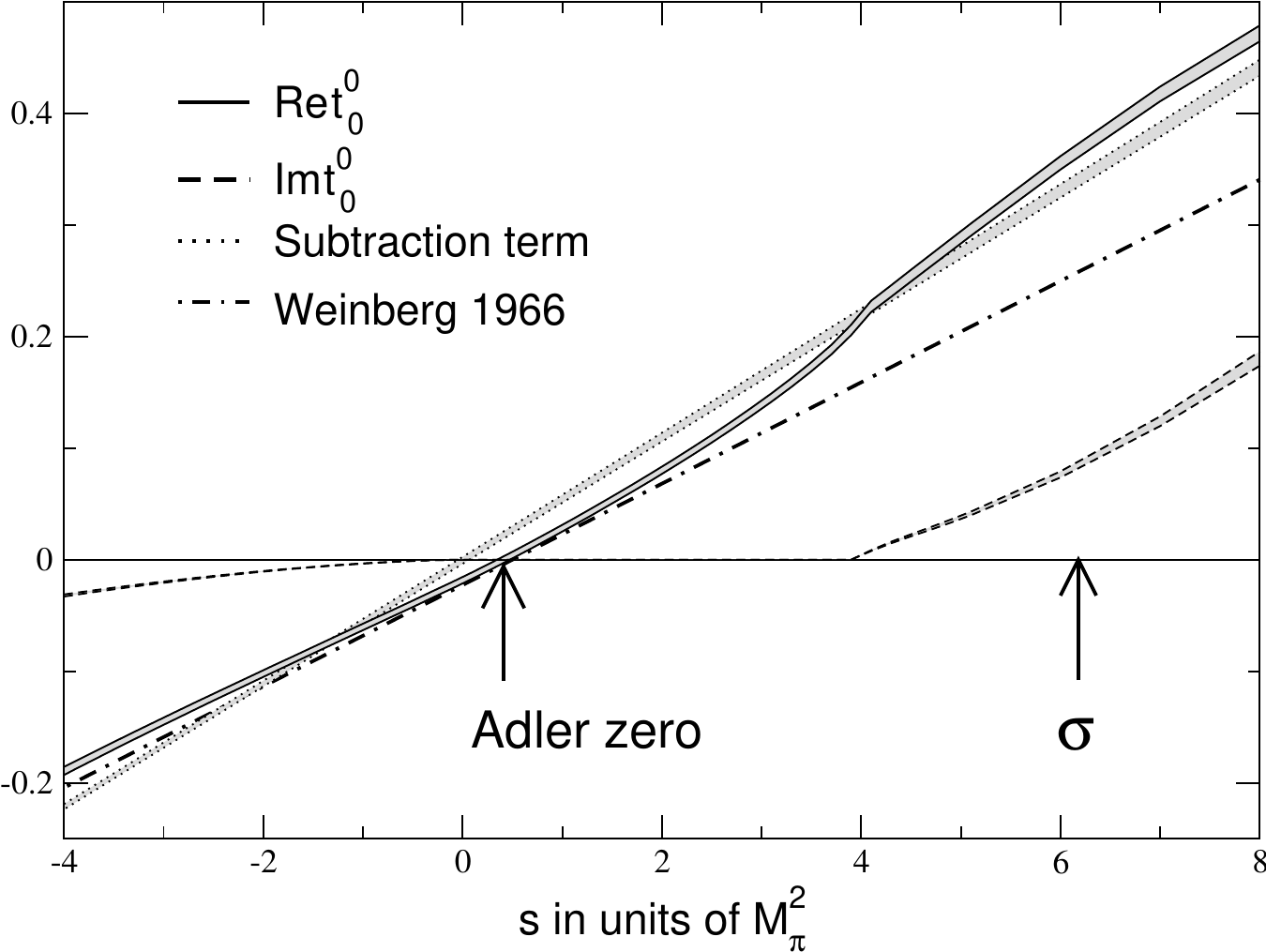}
\includegraphics[width=6.4cm]{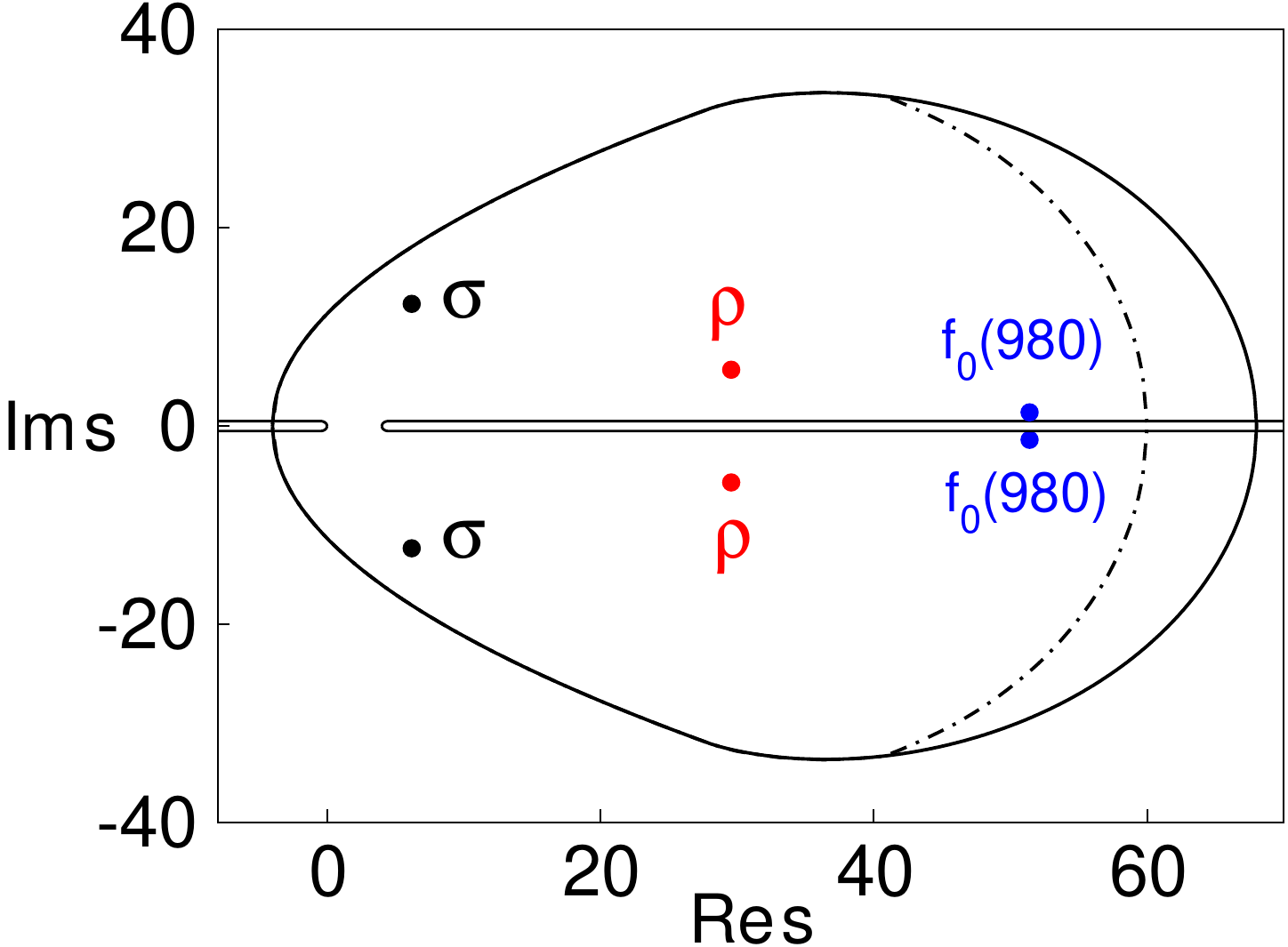}
\caption{ Left: the $S0$ partial wave on the real axis in the threshold
  region, showing the contribution of the subtraction polynomial vs. the
  rest. Right: position of the $\sigma$, the $\rho$ and the $f_0(980)$
  poles in the complex plane as determined from the solutions of the Roy
  equations. The contour indicates the limit of validity of these
  equations, the dashed line the one originally derived by Roy, and the
  solid line, the extended version thereof which was used in the analysis
  discussed here. Figures from Ref.~\cite{Caprini:2005zr}.
\label{fig:sigma}}
\end{figure}
The details of the analysis can be found in Ref.~\cite{Caprini:2005zr},
here I only present the final numerical result\footnote{I have eliminated
  the asymmetric uncertainty in the original paper, because this has been
  superseded in the meanwhile, see~\cite{Caprini:2011ky}.}
\be
M_\sigma=441 \pm 8 \, \mathrm{MeV}\; , \qquad \Gamma_\sigma=544 \pm 18 \,
\mathrm{MeV} \; ,
\ee
which represents a stunningly precise result for a resonance whose
existence has often been questioned (as witnessed by its exclusion in
several editions of the PDG). The reason for this precision can be
understood by looking at Fig.~\ref{fig:sigma}, left panel, which shows that
the subtraction polynomial is the most important contribution in the
threshold region, which is where the $\sigma$ pole lies: $s_\sigma=(6.2 \pm
i 12.3)M_\pi^2$. So, it is the precise knowledge of the scattering lengths
which is responsible for the precision seen in the $\sigma$ pole
determination. A picture of the position of this pole (together with those
of the $\rho$ and the $f_0(980)$) on the complex plane is shown on the
right panel of Fig.~\ref{fig:sigma}.

\section{Final remarks}
Dispersion relations in quantum field theory is a vast subject, to which
several books are dedicated, but which were written mainly in the sixties
and seventies. In the last two decades or so, this subject is now met with
renewed interest, in particular in view of the possibility to use it in
combination with effective field theories. The two approaches perfectly
complement each other and their combined use can lead to results of
remarkable precision, as I have tried to illustrate here.

In the limited time (for the lectures) and space (for these notes) which I
had available I could only provide a brief introduction to this fascinating
subject. I have chosen to touch on a few topics of current phenomenological
interest (foremost $(g-2)_\mu$, but also the pion vector form factor and
$\pi \pi$ scattering), and to follow a logical path which starts from the
two-point function, moves to the three-point function and then to the
four-point function. The two-point function needs as input the three-point
one, which needs as input the four-point one. If one only considers elastic
processes, the path stops there, because the four-point function is related
to itself by unitarity. This leads to nonlinear integral equations, whose
mathematics is very interesting and probably only partly understood, at
least for what concerns an efficient numerical treatment. The numerical
solutions of these equations show how powerful dispersion relations can be,
since they produce physical results out of pure mathematics (well, not
quite, since one needs some physics input too, but still).

As final remarks I wish to mention two important applications which have
been left out, not only because of lack of space, but also because the
dispersion relations themselves become quite intricate and are much less
suited for a pedagogical introduction. But I find it particularly important
to mention them here, to show to interested students that this is a
subject which can still offer difficult challenges and significant rewards
for those who are able to solve them.

The first is $\eta \to 3 \pi$. At first sight this doesn't look much more
complicated than $\pi \pi$ scattering itself: it is still a four-point
function, with one external particle heavier than the others, which allows
it to decay into the other three. The fact that the heavy particle is part
of the meson octet and forces one to consider SU(3) \chpt instead of SU(2)
is indeed a complication, but affects mainly the \chpt side of the
problem. For what concerns the dispersive approach, one still has to
consider mainly (well, only) pion rescattering effects, since
thresholds involving mesons with strangeness are high enough to be booked
as inelastic effects, which can be taken into account separately. Despite
these similarities, the complexity of the dispersion relation is
significantly higher than that of the Roy equations. This was discussed
first by Khuri and Treiman in 1960 for the $ K \to 3 \pi$
decay~\cite{Khuri:1960zz}, then taken up again in the nineties to combine
the dispersive approach with \chpt~\cite{Kambor:1995yc,Anisovich:1996tx}.
But a significant simplification in solving the relevant integral equations
was first discovered only recently by Gasser and
Rusetsky~\cite{Gasser:2018qtg}. Their method was adopted in a recent
comprehensive analysis of this decay, carried out by S.~Lanz, H.~Leutwyler,
E.~Passemar and myself~\cite{Colangelo:2016jmc,Colangelo:2018jxw}. The
reward in this case is a unique and reliable phenomenological source of
information on the $m_u-m_d$ quark mass difference. This is particularly
important, because otherwise we would only have lattice-based
determinations on the masses of up and down quarks, see the discussion in
the FLAG review~\cite{FlavourLatticeAveragingGroupFLAG:2021npn} and
in~\cite{Colangelo:2018jxw}.

The second one is the hadronic light-by-light (HLbL) contribution to
$a_\mu$. In contrast to the HVP, which has been treated with a dispersive
approach since the sixties, a dispersive treatment for this four-point
function has been deemed impossible for many years. The breakthrough was
achieved only about ten years ago by M.~Hoferichter, M.~Procura, P.~Stoffer
and myself. In a series of papers, we laid out the relevant formalism and
derived a master formula for the evaluation of this contribution to
$a_\mu$~\cite{Colangelo:2014dfa,Colangelo:2015ama,Colangelo:2017fiz,Colangelo:2017qdm}.
An improved evaluation of this contribution based on this formalism was
adopted in the first White Paper of the Muon $g-2$ Theory
Initiative~\cite{Aoyama:2020ynm}, and made it a subdominant source of
uncertainty. But a full exploitation of the power of the dispersive
approach required to solve significant and conceptual challenges in the
treatment of some intermediate states. These have been
solved only very recently by Hoferichter, Stoffer and
Zillinger~\cite{Hoferichter:2024fsj} who were also able to perform a
phenomenological analysis and provide the first completely dispersive
evaluation of this
contribution~\cite{Hoferichter:2024bae,Hoferichter:2024vbu}. This
evaluation, together with those made on the lattice, forms the basis for
the estimate provided in the second White Paper~\cite{Aliberti:2025beg}.
This is perhaps the best example of a very challenging application of
dispersion relations, with extremely complex and subtle theoretical
difficulties, which, once solved, bring a very significant reward. At the
current level of precision, the HLbL contribution, which only fifteen years
ago was considered a stumbling block for the Muon $g-2$, is now a solved
problem.

%\vskip 1.5cm

\bmhead{Acknowledgements} These lectures were given at the Summer School
``Continuum Foundation of Lattice Gauge Theories'', which took place at
CERN, July 22-26 2024. I thank the organizers of the School for the
invitation to be part of a stimulating set of lectures on interesting
subjects and to the students for their active participation. I very much
enjoyed not only giving my lectures, but also listening to those of my
fellow lecturers. In the spring semester 2025 I gave a graduate course on
the same subject and with about the same content at the University of Bern
and wish to thank all the students who attended the lectures, in particular the
group coming every week from PSI, for their engagement.  A special thanks
to Martina Cottini, J\"urg Gasser and Heiri Leutwyler for a careful reading
of the manuscript and their very useful comments.

\newpage

%\vskip 1.5 cm

\bibliography{DRL}

\end{document}